\newif\ifpreprint
\preprintfalse
\preprinttrue

\ifpreprint
	\documentstyle[equations,fleqn,epsf]{mn}
\else
	\documentstyle[equations,fleqn,epsf,referee]{mn}
\fi

\newif\ifAMStwofonts

\ifpreprint
\fi

\newcommand{\bnabla}	{{\bmath{\nabla}}}

\newcommand{\bv}	{{\bmath{v}}}
\newcommand{\br}	{{\bmath{r}}}
\newcommand{\bx}	{{\bmath{x}}}
\newcommand{\by}	{{\bmath{y}}}
\newcommand{\bz}	{{\bmath{z}}}
\newcommand{\bF}	{{\bmath{F}}}
\newcommand{\dx}	{\!{\rm d}^3\!\bmath{x}}
\newcommand{\dy}	{\!{\rm d}^3\!\bmath{y}}
\newcommand{\dz}	{\!{\rm d}^3\!\bmath{z}}
\newcommand{\dv}	{\!{\rm d}^3\!\bmath{v}}
\newcommand{\hn}	{\hat{n}}
\newcommand{\hrho}	{\hat{\rho}}
\newcommand{\hPhi}	{\hat{\Phi}}
\newcommand{\hbF}	{\hat{\bF}}
\newcommand{\Nbd}	{\mbox{$N$-body}}
\newcommand{\miseF}	{\mbox{${\rm MISE}(\hbF)$}}
\newcommand{\moptF}	{\mbox{${\rm MISE}_{\rm opt}(\hbF)$}}
\newcommand{\isbF}	{\mbox{${\rm ISB}(\hbF)$}}
\newcommand{\ivF}	{\mbox{${\rm IV}(\hbF)$}}

\newcommand{\eqn}[1]	{equation (\ref{#1})}
\newcommand{\Eqns}[1]	{Equations (\ref{#1})}
\newcommand{\eqns}[1]	{equations (\ref{#1})}
\newcommand{\eqi}[1]	{(equation \ref{#1})}
\newcommand{\eqis}[1]	{(equations \ref{#1})}
\newcommand{\eqb}[1]	{(\ref{#1})}

\ifoldfss
  \ifCUPmtlplainloaded \else
    \NewTextAlphabet{textbfit} {cmbxti10} {}
    \NewTextAlphabet{textbfss} {cmssbx10} {}
    \NewMathAlphabet{mathbfit} {cmbxti10} {} 
    \NewMathAlphabet{mathbfss} {cmssbx10} {} 
  \fi
  \ifAMStwofonts
    \ifCUPmtlplainloaded \else
      \NewSymbolFont{upmath} {eurm10}
      \NewSymbolFont{AMSa} {msam10}
      \NewMathSymbol{\upi}     {0}{upmath}{19}
      \NewMathSymbol{\umu}     {0}{upmath}{16}
      \NewMathSymbol{\upartial}{0}{upmath}{40}
      \NewMathSymbol{\leqslant}{3}{AMSa}{36}
      \NewMathSymbol{\geqslant}{3}{AMSa}{3E}

       \let\le=\leqslant
       \let\ge=\geqslant
    \fi
  \fi
\fi 

\ifnfssone
  \newmathalphabet{\mathit}
  \addtoversion{normal}{\mathit}{cmr}{m}{it}
  \addtoversion{bold}{\mathit}{cmr}{bx}{it}
  \newmathalphabet{\mathbfit} 
  \addtoversion{normal}{\mathbfit}{cmr}{bx}{it}
  \addtoversion{bold}{\mathbfit}{cmr}{bx}{it}
  \newmathalphabet{\mathbfss} 
  \addtoversion{normal}{\mathbfss}{cmss}{bx}{n}
  \addtoversion{bold}{\mathbfss}{cmss}{bx}{n}
  \ifAMStwofonts
    \ifCUPmtlplainloaded \else
      \UseAMStwoboldmath
      \makeatletter
      \new@mathgroup\upmath@group
      \define@mathgroup\mv@normal\upmath@group{eur}{m}{n}
      \define@mathgroup\mv@bold\upmath@group{eur}{b}{n}
      \edef\UPM{\hexnumber\upmath@group}
      \new@mathgroup\amsa@group
      \define@mathgroup\mv@normal\amsa@group{msa}{m}{n}
      \define@mathgroup\mv@bold\amsa@group{msa}{m}{n}
      \edef\AMSa{\hexnumber\amsa@group}
      \makeatother
      \mathchardef\upi="0\UPM19
      \mathchardef\umu="0\UPM16
      \mathchardef\upartial="0\UPM40
      \mathchardef\leqslant="3\AMSa36
      \mathchardef\geqslant="3\AMSa3E

       \let\le=\leqslant
       \let\ge=\geqslant
    \fi
  \fi
\fi 

\ifnfsstwo
  \DeclareMathAlphabet{\mathbfit}{OT1}{cmr}{bx}{it}
  \SetMathAlphabet\mathbfit{bold}{OT1}{cmr}{bx}{it}
  \DeclareMathAlphabet{\mathbfss}{OT1}{cmss}{bx}{n}
  \SetMathAlphabet\mathbfss{bold}{OT1}{cmss}{bx}{n}
  \ifAMStwofonts
    \ifCUPmtlplainloaded \else
      \DeclareSymbolFont{UPM}{U}{eur}{m}{n}
      \SetSymbolFont{UPM}{bold}{U}{eur}{b}{n}
      \DeclareSymbolFont{AMSa}{U}{msa}{m}{n}
      \DeclareMathSymbol{\upi}{0}{UPM}{"19}
      \DeclareMathSymbol{\umu}{0}{UPM}{"16}
      \DeclareMathSymbol{\upartial}{0}{UPM}{"40}
      \DeclareMathSymbol{\leqslant}{3}{AMSa}{"36}
      \DeclareMathSymbol{\geqslant}{3}{AMSa}{"3E}

       \let\le=\leqslant
       \let\ge=\geqslant
    \fi
  \fi
\fi 

\ifCUPmtlplainloaded \else
  \ifAMStwofonts \else 
    \def\upi{\pi}
    \def\umu{\mu}
    \def\upartial{\partial}
  \fi
\fi

\title[Towards optimal softening in 3D \Nbd\ codes: I. Minimizing the force error]
        {Towards optimal softening in 3D $\bmath{N}$-body codes: \\
        {\LARGE\bf I. Minimizing the force error}}
\author[Walter Dehnen]
        {Walter Dehnen \\
         Max-Planck Institut f\"ur Astronomie, 
         K\"onigstuhl 17, D69117 Heidelberg, Germany; 
         dehnen@mpia-hd.mpg.de}

\begin{document}

\maketitle

\begin{abstract}
        In \Nbd\ simulations of collisionless stellar systems, the forces are
        softened to reduce the large fluctuations due to shot noise. Softening
        essentially modifies the law of gravity at $r=|\bx_i-\bx_j|$ smaller
        than some softening length $\epsilon$. Therefore, the softened forces
        are increasingly biased for ever larger $\epsilon$, and there is some
        optimal $\epsilon$ which yields the best compromise between reducing the
        fluctuations and introducing a bias. Here, analytical relations are
        derived for the amplitudes of the bias and the fluctuations in the limit
        of $\epsilon$ being much smaller than any physical scale of the
        underlying stellar system. In particular, it is shown that the
        fluctuations of the force are generated locally, in contrast to the
        variations of the potential, which originate from noise in the whole
        system. Based on these asymptotic relations and using numerical
        simulations, I study the dependence of the resulting force error on the
        number of bodies, the softening length, and on the functional form by
        which Newtonian gravity is replaced. The widely used Plummer softening,
        where each body is replaced by a Plummer sphere of scale radius
        $\epsilon$, yields significantly larger force errors than do methods in
        which the bodies are replaced by density kernels of finite extent. I
        also give special kernels, which reduce the errors even further. These
        kernels largely compensate the errors made with too small inter-particle
        forces at $r<\epsilon$ by exceeding Newtonian forces at $r\sim\epsilon$.
        Additionally, the possibilities of locally adapting $\epsilon$ and of
        using unequal weights for the bodies are investigated. By using these
        various techniques without increasing $N$, the rms force error can be
        reduced by a factor $\sim2$ when compared to the standard Plummer
        softening with constant $\epsilon$. The results of this study are
        directly relevant to \Nbd\ simulations using direct summation techniques
        or the tree code for force evaluation.
\end{abstract}

\begin{keywords}
        \Nbd\ simulations ---
        stellar dynamics --- methods: numerical
\end{keywords}

\section{Introduction}\label{sec:intro}
\Nbd\ techniques are applied to stellar dynamical problems ranging from
collision-dominated systems, like stellar clusters and galactic centres, to
systems with extremely large numbers of particles, such as galaxies and
large-scale structure. In the first case, one is actually interested in
simulating a system of $N$ gravitationally interacting particles, which is done
by {\em collisional\/} \Nbd\ codes.

In systems with very large numbers of particles, on the other hand, the
two-particle relaxation time greatly exceeds the age, i.e.\ these systems behave
essentially {\em collisionless,} very much like a continuous system. In other
words, the shot noise is negligible, and each particle essentially moves in the
mean force field generated by all other particles.  Thus, the aim of any
collisionless \Nbd\ simulation is to simultaneously solve the collisionless
Boltzmann equation (CBE)
\begin{equation} \label{cbe}
        {\partial f\over\partial t} + {\partial f\over\partial\bx}\cdot\bv
                        + {\partial f\over\partial\bv}\cdot\bF = 0,
\end{equation}
describing the stellar dynamics, and
\begin{equation} \label{grav}
	\bF(\bx,t) = -G \int\dx^\prime\;\int\dv\;
		{\bx-\bx^\prime\over|\bx-\bx^\prime|^3}\,f(\bx^\prime,\bv,t)
\end{equation}
describing gravity. Here, $f(\bx,\bv,t)$ is the (continuous) one-particle%
	\footnote{Correlations between particles (e.g.\ binaries) are not
	described by $f$ but by the two-particle distribution function and
	higher orders in the BBGKY hierarchy (see, e.g., Binney \& Tremaine
	1987).}
phase-space distribution function (DF) of the stellar system, while $\bF(\bx,t)$
is the gravitational force field generated by it. A numerical treatment of the
CBE on an Eulerian grid is not feasible because of the vast size of
six-dimensional phase space and the strong inhomogeneity of the DF. Instead, one
applies the method of characteristics. Since the CBE states that $f$ is constant
along any trajectory $\{\bx(t),\bv(t)\}$, the trajectories obtained by
time-integration of $N$ points $\{\bx_i,\bv_i\}$ sampled from the DF at $t=0$
form a representative sample of $f$ at each time $t$.

The result of this procedure is a \Nbd\ code, however, with a crucial difference
to collisional \Nbd\ codes: the bodies do not represent real particles, rather
they are a Monte-Carlo representation of the underlying continuous DF. In fact,
all the information on $f(\bx,\bv,t)$ is given by the phase-space positions of
the bodies. Since $N$ is much smaller than the number of particles in the system
being modelled, this information is always incomplete.

\subsection{The rationale of force softening} \label{sec:intro:why}
Having solved for the dynamics, we still need to compute gravity. The
simple-minded approach of replacing $f$ in \eqn{grav} with a sum of
$\delta$-peaks at the body positions corresponds to a Monte-Carlo integration
\cite{pp93} and yields the forces of $N$ mutually interacting particles.

In practice, one {\em softens\/} the forces at small inter-body separations
$r=|\bx^\prime-\bx|$, which is equivalent to replacing the bodies by some extended
mass distribution instead of $\delta$-functions\footnote{
	In describing their dynamics, on the other hand, the bodies are treated
	as {\em point-like\/} test particles. Dyer \& Ip (1993) argued that this
	asymmetry is physically inconsistent. However, bodies do not represent
	physical particles and no physical inconsistency can arise. The goal is
	{\em not\/} the simulation of $N$ mutually interacting particles
	point-like or not.}.
Since there appears to be some confusion for the proper reasons to introduce
softening, I will now try to give a thorough motivation.

\subsubsection{Suppressing artificial relaxation? No!} \label{sec:intro:relax}
Contrary to widespread belief, softening does not much reduce the (artificial)
two-body relaxation \cite{theis98}, the time-scale of which scales like $N/\ln
N$ and which cannot be neglected in {\em simulations\/} of collisionless stellar
systems.

To see this, recall the works of Chandrasekhar and Spitzer, who showed that
two-body relaxation is driven by close as well as distant encounters: each
octave in distance is contributing equally. By softening the forces at small
$r$, one reduces the contributions from close encounters. Most of the
relaxation, however, is due to noise on larger scales, and hence cannot be
avoided by softening techniques, in agreement with the numerical findings of
Hernquist \& Barnes (1990).
\subsubsection{Avoiding artificial two-body encounters} \label{sec:intro:avoid}
Without softening, the accurate numerical integration of close encounters and
binaries, the main obstacle in collisional \Nbd\ codes, requires great care and
substantial amounts of computer time. However, as already mentioned, the bodies
are just a representation of the {\em one-particle\/} DF, and consequently
binaries as well as two-body encounters (and the Newtonian forces arising in
them) are entirely artificial. Thus, softening may be motivated by the desire to
suppress such artifacts.
\subsubsection{Estimating the true forces} \label{sec:intro:estim}
One may consider force softening as a way to obtain, at each time step, from the
$N$ body positions an {\em estimate\/} for the gravitational forces generated by
the continuous mass density of the system being modelled (Merritt 1996, see also
below). In this context, softening essentially implies the assumption that the
mass density is smooth on small scales. This assumption compensates for the
aforementioned incompleteness in our knowledge of the DF, a standard method in
non-parametric estimation techniques.
\subsubsection{The benefits of softening}
As a benefit of softening, close encounters between bodies are much less of a
problem: one can use the simple leap-frog integrator, a method that would be
highly inefficient in collisional \Nbd\ codes. Moreover, since only an estimated
force fields is computed, approximate, and hence cheaper, methods for its
computation than direct summation may be used, as long as the errors introduced
by the approximation are smaller than those due to the estimation itself. Both
these effects lead to substantial savings of computer time, which in turn enable
much larger numbers of bodies in collisionless than in collisional \Nbd\ codes
(currently about 3-4 orders of magnitude more). This may be considered the real
motivation for softening.

\subsection{Optimal softening}
Softening also has a severe drawback: it artificially modifies mutual
gravitational forces at small inter-body separations: instead of Newton's
inverse square law, the forces vanish in the limit $r\to0$ (necessary to yield a
continuous $\bF(\bx)$). With increasing degree of softening, this results in an
ever stronger {\em bias\/}, the systematic error of the mean estimated as
compared to the true forces, corresponding to a diminution of the information
hold in the body positions.

Thus, softening allows a trade-off between artificially high near-neighbour
forces and a biased force field. Both do affect the fidelity but also the
efficiency of a \Nbd\ simulation. With little softening, the integration of the
equations of motion requires great care, demanding substantial computer
resources (if the time integration is not done carefully enough, two-body
relaxation may be strongly enhanced and other unwanted effects may arise). On
the other hand, the force bias introduced by softening modifies the dynamics,
possibly on scales much larger than that on which forces are softened, which
might deteriorate the simulation. The influence of these two effects on a \Nbd\
simulation will certainly depend on (i) the purpose of the simulation, (ii) the
properties of the system modelled and (iii) the time span integrated. This
implies that the optimal way of softening depends on the nature of the problem
investigated as well as the goal of the simulation.

The idea of this paper is to make a first step towards answering the question of
the optimal softening by investigating the effect of softening on the estimated
force. That is, here we consider only the {\em static\/} problem of optimal
force estimation. An investigation of the effect of softening on the {\em
dynamics\/} is beyond the scope of this paper, but see the discussion in
Section~\ref{sec:disc:eopt}.

In some situations, one can reduce the noise by a careful rather than random
choice of the trajectories, the so-called ``quiet start'' technique
\cite{se83}. However, if chaotic motion dominates most of the trajectories,
e.g.\ in a merger simulation, the benefits of such a careful set-up quickly
disappear. Here, I am interested in the general, i.e.\ worst, case, and assume
throughout that the trajectories are sampled randomly from the DF.

\subsection{A formalism of softening}
A softening technique that is widely used is the Plummer softening, initiated by
Aarseth's (1963) use of Plummer spheres to model galaxies in a cluster. In this
method, each body is replaced by a Plummer sphere with scale radius $\epsilon$,
resulting in the potential estimator (hereafter, an estimate for some quantity
is indicated by a hat)
\begin{equation} \label{plummer-soft}
        \hPhi(\bx) =  -G \sum_{i=1}^N {m\over\sqrt{\epsilon^2+(\bx-\bx_i)^2}},
\end{equation}
where $m=M/N$ is the weight (or mass) of each individual body. A more general
estimator for the gravitational potential is
\begin{subequations} \label{kern}
\begin{equation} \label{soft:pot}
        \hPhi(\bx) = -G \sum_{i=1}^N {m\over\epsilon}\;
                \phi\left[{|\bx-\bx_i|\over\epsilon}\right].
\end{equation}
This formula clearly separates the two aspects of the softening method: the {\em
softening kernel\/} $\phi(r)$, which determines the functional form of the
modified gravity, and the {\em softening length\/} $\epsilon$. The Plummer
softening, for instance, corresponds to $\phi=(1+r^2)^{-1/2}$. The corresponding
estimates for the force field $\bF=-\bnabla\Phi$ and the density $\rho$ of the
underlying smooth system are
\begin{eqnarray} \label{soft:force}
        \hbF(\bx) &=&-G \sum_{i=1}^N {m\over\epsilon^2}\,
                \phi^\prime\left[{|\bx-\bx_i|\over\epsilon}\right]\,
                {\bx-\bx_i\over|\bx-\bx_i|},                    \\
\label{soft:density}
        \hrho(\bx) &=& \sum_{i=1}^N {m\over\epsilon^3}\,
                \eta\left[{|\bx-\bx_i|\over\epsilon}\right],
\end{eqnarray} \end{subequations}
where
\begin{equation}
	\eta(r) =  -{1\over4\pi r^2}\, {\partial\over\partial r} \Big(r^2\,
 		{\partial\phi\over\partial r}\Big),
\end{equation} 
is the {\em kernel density}, the mass distribution by which each body is
replaced in these estimates. Note that kernels $\phi(r)$ which reproduce
Newtonian gravity exactly for $r$ larger than some finite radius $r_0$
correspond to density kernels of finite support, i.e.\ $\eta(r)\equiv0$ for
$r>r_0$.

Following the terminology of statisticians (e.g..\ Silverman 1986), the above
approximations in \eqns{kern} may be called {\em fixed-kernel estimators\/}
because the softening length $\epsilon$ is the same for all bodies.

\subsection{Outline of the paper}
The reader might be surprised by this introduction being rather void of
quotations. In fact, even though \Nbd\ methods with softening are widely used by
astronomers, no detailled investigation of the properties of force softening has
yet been undertaken. Some papers deal with the small-scale truncation of the
mutual forces, but only few consider this bias in conjunction with the reduction
of noise. Merritt \& Tremblay (1994) dealt with estimators for the density of
stellar systems, and pointed out the close connection to softening in \Nbd\
simulations. Merritt (1996) introduced the important concept of the mean square
force error (see Section~\ref{sec:errors} below) and studied empirically the
effect of the softening length on the errors of the forces estimated by Plummer
softening.  Athanassoula et~al.\ (1998, 2000) have extended this work to larger
$N$ and to using the tree code rather than direct summation as Poisson solver.

A systematic investigation of the dependence of the force errors on the
softening method, in particular for other than Plummer softening, is still
missing, and the goal of this paper is to fill this gap. Merritt and
Athanassoula et~al.\ found, by brute force, empirical relations for the
dependence on $N$ of the optimal softening length, $\epsilon_{\rm opt}$, and the
resulting force error, mainly for the case of Plummer softening. Instead, I
derive in Section~\ref{sec:errors} and Appendix~\ref{app:asym} of this paper
analytic expressions, valid for softening of the form \eqb{kern}, for the force
error in the asymptotic limit of small $\epsilon$ and large $N$. These
asymptotic relations apply whenever $\epsilon$ is smaller than any physical
scale of the stellar system being modelled, a condition that must be satisfied
for the \Nbd\ simulation to be of any use. Based on the results of this
investigation, I discuss in Section~\ref{sec:opt} the optimal softening length
and kernel, and the asymptotic scaling with $N$ of $\epsilon_{\rm opt}$ and the
error achieved, while numerical experiments that demonstrate and verify these
analytic results are presented in Section~\ref{sec:exper}.

Apart from just increasing the number $N$ of bodies and of the choice of the
softening kernel and length, the \Nbd\ experimenter may use a scheme to locally
adapt the softening, and use different individual weights. These two latter
techniques, which aim at enhancing the resolution in regions of higher density
and thus decreasing the force error, are discussed in Section~\ref{sec:ind} and
related numerical experiments are presented in Section~\ref{sec:exper:adaptive}.
In Section~\ref{sec:disc}, I discuss the relevance of the results of these
investigations to other Poisson solvers than direct summation, and touch the
question for the optimal collisionless \Nbd\ code and related issues. Finally,
Section~\ref{sec:summ} sums up and concludes.
\section{The errors of the estimates} \label{sec:errors}
Let us consider the {\em mean square error\/} (MSE) of the estimate
$\hat{a}(\bx)$ for some field $a(\bx)$
\begin{equation} \label{mse}
        {\rm MSE}_\bx(\hat{a}) =
	\left\langle\big[\hat{a}(\bx)-a(\bx)\big]^2\right\rangle,
\end{equation}
where $\langle\rangle$ denotes the expectation value, the ensemble average over
all random realizations of the underlying smooth density distribution
$\rho(\bx)$ with $N$ positions. Elementary manipulations yield
\begin{equation} \label{m=b+v}
        {\rm MSE}_\bx(\hat{a}) 	= \big[{\rm bias}_\bx(\hat{a})\big]^2
				+ {\rm var}_\bx(\hat{a})
\end{equation}
with
\begin{eqnarray} 
\label{bias}{\rm bias}_\bx(\hat{a}) &=& \big\langle\hat{a}(\bx)\big\rangle
					- a(\bx), \\
\label{var} {\rm var}_\bx(\hat{a})  &=& \left\langle\big[\hat{a}(\bx)
                                 - \big\langle\hat{a}(\bx)\big\rangle
					\big]^2\right\rangle 
                                =  \left\langle\hat{a}^2(\bx)\right\rangle
                                 - \big\langle\hat{a}(\bx)\big\rangle^2.
\end{eqnarray}
When applied to our problem, the bias is the deviation of the softened potential
and forces in the continuum limit $N\to\infty$ from the true potential and
forces of the underlying system. The variance, on the other hand, describes the
amplitude of the random deviations of the estimated potential and force from
their expectation value. Softening reduces the variance but introduces a
bias. Thus, there must be an optimal amount of softening between these two
extremes \cite{merritt96}.
\subsection{Asymptotic behaviour of the bias} \label{sec:asym:bias}
The behaviour of bias and variance for small $\epsilon$ is derived analytically
in Appendix~\ref{app:fixed:bias}. For the biases of the softened potential
\eqb{soft:pot}, force \eqb{soft:force}, and density \eqb{soft:density}, one finds
\begin{subequations} \label{biases} \begin{eqnarray} \label{bias:pot}
        {\rm bias}_\bx(\hPhi) &=&    \phantom{-}
				  a_0\,\epsilon^2G\,\rho(\bx)
                                + a_2\,\epsilon^4G\,\Delta\rho(\bx)
                                + {\cal O}(\epsilon^6),
\\[1ex] \label{bias:force}
        {\rm bias}_\bx(\hbF)  &=&
				- a_0\,\epsilon^2G\,\bnabla\!\rho(\bx)
                                - a_2\,\epsilon^4G\,\bnabla\!\Delta\rho(\bx)
                                + {\cal O}(\epsilon^6),
\\[1ex] \label{bias:density}
        {\rm bias}_\bx(\hrho) &=&  \phantom{-}
				  {a_0\over4\pi}\,\epsilon^2\Delta\rho(\bx)
                                + {a_2\over4\pi}\,\epsilon^4\Delta\Delta\rho(\bx)
                                + {\cal O}(\epsilon^6)
\end{eqnarray} \end{subequations}
(e.g.\ Hernquist \& Katz 1989). Here, $a_k$ are constants that depend only on the
kernel:
\begin{subequations} \label{ak} \begin{eqnarray} \label{as:phi}
        a_k     &=& {4\pi\over(k+1)!} \int_0^\infty{\rm d}r\;r^{k+2}
                        \left[{1\over r}-\phi(r)\right]
\\[1ex]\label{as:rho}           
                &=& {(4\pi)^2\over(k+3)!} \int_0^\infty{\rm d}r\;r^{k+4}\,
                        \eta(r).
\end{eqnarray} \end{subequations}
\Eqns{biases} are based on the low orders of a Talyor expansion of 
$\rho(\bx-\epsilon\bz)$ around $\bx$. Thus, if the contributions of the
higher-order terms are non-neglible within the softening volume, these equations
fail. Such a failure may occur for various reasons. First, if $\epsilon$ is not
much smaller than the local physical scale of the underlying system, the
softening volume is large enough for the higher-order terms to contribute.
Second, the underlying stellar system may have power on all physical scales, for
instance, in a density cusp. In such a case, the Taylor expansion does not
converge, but the bias is usually smaller than \eqns{biases} predict (see
\eqns{bias:cusp}).

Finally, \eqns{biases} fail if the kernel density $\eta(r)$ approaches zero not
fast enough at large $r$. In this case, the integrals in \eqns{ak} do not exist,
and the bias (i) cannot be described by local quantities and (ii) grows faster
than $\epsilon^2$. Hereafter, kernels with existing $a_0$ will be called {\em
compact}. At large radii, the density of a compact kernel is either identical
zero or falls off more steeply than $r^{-5}$.

For compact kernels with positive definite $\eta(r)$, $0<a_0<\infty$ and these
biases have some notable properties. First, they increase only like
$\epsilon^2$.  Second, the bias of the potential is proportional to the density
and hence everywhere positive, i.e.\ gravity is always under-estimated. Third,
the biases are smaller for more compact kernels, as measured by $a_0$. And, of
course, the biases are independent of $N$, since they describe the deviation
from the smooth underlying model in the limit $N\to\infty$ (this last point
holds for any kernel).
\subsection{Asymptotic behaviour of the variance} \label{sec:asym:var}
Assuming the $N$ positions $\bx_i$ are independent, the variances of various
estimates are $N^{-1}$ times the variance due to the estimate obtained from just
one body, and, by virtue of the central-limit theorem, the distribution of the
estimates is nearly normal. For the asymptotic behaviour at small $\epsilon$ I
derive in Appendix~\ref{app:fixed:var}
\begin{subequations} \label{vars} \begin{eqnarray} \label{var:pot}
        N{\rm var}_\bx(\hPhi) &=& {\rm var}_\bx^0(\Phi)
                - b_\Phi\,G^2\,M\,\epsilon\,\rho(\bx)
                \;+\; {\cal O}(\epsilon^2),
\\[1ex] \label{var:force}
        N{\rm var}_\bx(\hbF) &=& b_F\,G^2\,M\,\epsilon^{-1}\rho(\bx)
                                 \;+\; {\cal O}(\epsilon^0),
\\[1ex] \label{var:density}
        N{\rm var}_\bx(\hrho)&=& b_\rho\,M\,\epsilon^{-3}\rho(\bx)
                                 \;+\; {\cal O}(\epsilon^{-1}),
\end{eqnarray} \end{subequations}
where only the trace of the force variance is given, which to lowest order is
proportional to the unit matrix. Here,
\begin{equation} \label{var:zero}
        {\rm var}_\bx^0(\Phi) = G^2M\int\dy\;{\rho(\by)
                        \over|\bx-\by|^2}\;-\;\Phi^2(x)
\end{equation}
is the potential's variance per body in the absence of any softening, while
the constants
\begin{subequations} \label{bs} \begin{eqnarray}
        b_\Phi &=& 4\pi    \int_0^\infty {\rm d}r\;\big[1-r^2\phi^2(r)\big],
\\[1ex] b_F  &=& 4\pi    \int_0^\infty {\rm d}r\;r^2\phi^{\prime2}(r)
                = (4\pi)^2 \int_0^\infty {\rm d}r\;r^2\eta(r)\,\phi(r),
\\[1ex] b_\rho &=& 4\pi    \int_0^\infty {\rm d}r\;r^2\eta^2(r)
\end{eqnarray} \end{subequations}
depend on the kernel only.

Even though \eqns{vars} are of paramount importance for the understanding of
potential, force and density estimation, it appears that only \eqn{var:density}
has been derived previously (e.g.\ Silverman 1986), because density estimation
is a widely applied and well-established technique. There is a fundamental
difference between the behaviours of the variance of the potential on the one
side and that of the force or density on the other side. In the limit of
vanishing softening, the variance of the potential becomes $N^{-1}{\rm
var}^0_\bx(\hPhi)$, which is a finite {\em global\/} measure, while the
variances of force and density diverge in the limit of $\epsilon\to0$ and depend
on the {\em local\/} density. As a consequence, many of the textbook results on
density estimation will carry over into force estimation -- though with changes
in detail due to the different dependence on $\epsilon$ as $\epsilon\to0$ -- and
hence be useful for the \Nbd\ field.
\section{The Optimal force estimate} \label{sec:opt}
\subsection{The mean integrated square error}
So far, we have considered the error of the estimates at one point $\bx$. In
order to assess the global error introduced, one usually averages the local mean
square error over the whole stellar system to obtain the MISE (mean integrated
square error, e.g.\ Silverman 1986, Merritt 1996)
\begin{subequations} \label{mise+isb:iv} \begin{equation} \label{mise} 
        {\rm MISE}(\hat{a})  = M^{-1}\,\int\dx\;\rho(\bx)\, {\rm MSE}_\bx(\hat{a}).
\end{equation}
This can be divided into the integrated squared bias (ISB) and the integrated
variance (IV): MISE=ISB+IV, where
\begin{eqnarray} \label{isb} 
        {\rm ISB}(\hat{a})   &=& M^{-1}\,\int\dx\;\rho(\bx)\, 
                                        \big[{\rm bias}_\bx(\hat{a})\big]^2,
\\ \label{iv} 
        {\rm IV}(\hat{a})    &=& M^{-1}\,\int\dx\;\rho(\bx)\, {\rm var}_\bx(\hat{a}).
\end{eqnarray} \end{subequations}
We now apply the asymptotic relations of Section~\ref{sec:errors} in order to
obtain expressions for the mean integrated force error, \miseF, in the
asymptotic limit of small $\epsilon$ and large $N$.
\subsection{The optimal softening length}\label{sec:opt:eps}
\ifpreprint
\begin{table*}
        \caption{Some 3D softening kernels and their properties
                \label{tab:kernels}}
        \footnotesize
	\begin{tabular}{llcccccccr}
        $\!\!$name                                          &
        \hspace{4em}{kernel density $\eta(r)$}              &
        \multicolumn{1}{c}{$r_{{\rm max}\,F}$}              &
        \multicolumn{1}{c}{$\phi^\prime_{\rm max}$}         &
        \multicolumn{1}{c}{$a_0$}                           &
        \multicolumn{1}{c}{$a_2$}                           &
        \multicolumn{1}{c}{$b_\Phi$}                        &
        \multicolumn{1}{c}{$b_F$}                           &
        \multicolumn{1}{c}{$b_\rho$}                        &
        \multicolumn{1}{c}{$E_F$}                           \\  \hline
        $\!\!$Plummer
                & $\displaystyle{3\over4\pi}{1\over(1+r^2)^{5/2}}$
		& 0.707107 & 0.384900
                & $\infty$
                & $\infty$
                & $2\pi^2$
                & $\displaystyle{3\pi^2\over4}$
                & $\displaystyle{45\over1024}$
                & \multicolumn{1}{c}{--}
\\[4ex]	$\!\!$cubic spline  
                & $\left\{\begin{array}{l} {1\over4\pi}(4-6r^2+3r^3)    \\[1ex]
                           {1\over4\pi}(2-r)^3          \\[1ex]
                            0   \end{array}\right.$ \hfill
                  $\begin{array}{r} r<1 \\[1ex] 1\le r<2 \\[1ex] r\ge2
		   \end{array}$
		& 0.828302 & 0.657817
                & $\displaystyle{3\pi\over5}$
                & $\displaystyle{17\pi\over60}$
                & $\displaystyle{1120789\pi\over450450}$
                & $\displaystyle{70016\pi\over17325}$
                & $\displaystyle{491\over1260\pi}$
                & 9.84
\\[6ex]	$\!\!$F$_0$
		& $\displaystyle{3\over4\pi}\,H(1-r^2)$
		& 1 & 1
                & $\displaystyle{2\pi\over5}$
                & $\displaystyle{\pi\over70}$
                & $\displaystyle{72\pi\over35}$
                & $\displaystyle{24\pi\over5}$
                & $\displaystyle{3\over4\pi}$
                & 9.60
\\[3ex]	$\!\!$F$_1$
                & $\displaystyle{15\over8\pi}(1-r^2)\,H(1-r^2)$
		& 0.745356 & 1.242260
                & $\displaystyle{2\pi\over7}$ 
                & $\displaystyle{\pi\over126}$ 
                & $\displaystyle{400\pi\over231}$
                & $\displaystyle{40\pi\over7}$
                & $\displaystyle{15\over14\pi}$
                & 9.65
\\[3ex]$\!\!$F$_2$
                & $\displaystyle{105\over32\pi}(1-r^2)^2\,H(1-r^2)$
		& 0.592614 & 1.637096
                & $\displaystyle{2\pi\over9}$ 
                & $\displaystyle{\pi\over198}$ 
                & $\displaystyle{1960\pi\over1287}$
                & $\displaystyle{2800\pi\over429}$
                & $\displaystyle{35\over22\pi}$
                & 9.71
\\[3ex]$\!\!$F$_3$
                & $\displaystyle{315\over64\pi}(1-r^2)^3\,H(1-r^2)$
		& 0.505871 & 2.051564
                & $\displaystyle{2\pi\over11}$ 
                & $\displaystyle{\pi\over186}$ 
                & $\displaystyle{63504\pi\over46189}$
                & $\displaystyle{17640\pi\over2431}$
                & $\displaystyle{315\over143\pi}$
                & 9.75
\\[4ex] $\!\!$K$_0$   
                & $\displaystyle{15\over16\pi}(5-7r^2)\,H(1-r^2)$
		& 0.629941 & 2.624753
                & 0
                & $\displaystyle-{\pi\over126}$
                & $\displaystyle{170\pi\over231}$
                & $\displaystyle{10\pi}$
                & $\displaystyle{75\over16\pi}$
                & 9.43
\\[3ex]$\!\!$K$_1$   
                & $\displaystyle{105\over64\pi}(5-9r^2)(1-r^2)\,H(1-r^2)$
		& 0.493924 & 3.436176
                & 0
                & $\displaystyle-{\pi\over198}$ 
                & $\displaystyle{280\pi\over429}$
                & $\displaystyle{1610\pi\over143}$
                & $\displaystyle{525\over88\pi}$
                & 9.48
\\[3ex]$\!\!$K$_2$   
                & $\displaystyle{315\over128\pi}(5\,{-}\,11r^2)(1\,{-}\,r^2)^2
				 \,H(1-r^2)$
		& 0.419491 & 4.296037
                & 0
                & $\displaystyle-{\pi\over286}$ 
                & $\displaystyle{27342\pi\over46189}$
                & $\displaystyle{30240\pi\over2431}$
                & $\displaystyle{9135\over1144\pi}$
                & 9.54
\\[3ex]$\!\!$K$_3$   
                & $\displaystyle{3465\over1024\pi}(5\,{-}\,13r^2)(1\,{-}\,r^2)^3
			\,H(1-r^2)$
		& 0.370935 & 5.170685
                & 0
                & $\displaystyle-{\pi\over390}$ 
                & $\displaystyle{2772\pi\over5083}$
                & $\displaystyle{56826\pi\over4199}$
                & $\displaystyle{86625\over8398\pi}$
                & 9.60
\\ \hline
\end{tabular}\par\medskip\begin{minipage}{175mm}
	The kernels F$_n$ (Ferrers (1877) spheres of index $n$) and K$_n$ are
	continuous in the first $n$ force derivatives ($H$ denotes the Heaviside
	function). The potential of the Plummer kernel is $\phi(r)=1/
	\sqrt{1+r^2}$, that of the cubic spline kernel is given by Hernquist \&
	Katz (1989), while for the kernels F$_n$ and K$_n$, the potentials are
	low-order polynomials in $r^2$ and can be found in
	Appendix~\ref{app:kernels}.  $r_{{\rm max}\,F}$ is the radius, in units
	of $\epsilon$, of the maximal force, while $\phi^\prime_{\rm
	max}\equiv\phi^\prime( r_{{\rm max}\,F})$. The constants $a_0$ and $a_2$
	are related to the bias introduced by the softening \eqis{bias}, while
	the constants $b_\Phi$, $b_F$, and $b_\rho$ are related to the variances
	\eqis{vars}.  $E_F$ (defined in the text) is a measure for the
	efficiency of the kernel: for fixed $N$ and optimal $\epsilon$, the mean
	integrated squared force error, \miseF, is directly proportional to
	$E_F$, though with a different constant of proportionality for compact
	non-negative kernels (spline and F$_n$) and compensating kernels
	(K$_n$), respectively.\end{minipage}
\end{table*}
\fi
\subsubsection{Compact kernels with $a_0\neq0$}\label{sec:opt:nonneg}
For compact kernels with $\eta\ge0$, we have $a_0\neq0$ and inserting the
estimates \eqb{bias:force} and \eqb{var:force} into \eqns{mise+isb:iv} gives
\begin{equation} \label{miseF:asym}
        \miseF \approx A\,a_0^2\,\epsilon^4 + B\,b_F\,N^{-1}\epsilon^{-1}
\end{equation}
with
\begin{subequations} \begin{eqnarray}
\label{A} A&=&\textstyle G^2M^{-1}\int\dx\;\rho(\bx)
				  \big[\bnabla\rho(\bx)\big]^2,\\
\label{B} B&=&\textstyle G^2      \int\dx\;\rho^2(\bx),
\end{eqnarray} \end{subequations}
which depend on the underlying system. In the asymptotic relation
\eqb{miseF:asym} the dependences of $\miseF$ on the number $N$ of bodies, the
softening length $\epsilon$, the kernel function, and the underlying stellar
system are nicely separated. Choosing $\epsilon$ such as to minimize \miseF\
gives
\begin{subequations} \label{force:opt} \begin{eqnarray} \label{eps:opt}
        \epsilon_{\rm opt} &\approx& 4^{-1/5}\,(B/A)^{1/5}\,
                        (b_F/a_0^2)^{1/5}\,N^{-1/5},
\\[1ex] \label{mise:opt}
        \moptF &\approx& 4^{-4/5}5\,(A\,B^4)^{1/5}\,
                        (b_F^4\,a_0^2)^{1/5}\,N^{-4/5}.
\end{eqnarray} \end{subequations}
We also find that at optimal softening $\ivF=4\,\isbF$, i.e.\ the contributions
to the \miseF\ from variance and bias are of similar size. In the asymptotic
regime, the \miseF\ depends on the kernel through the combination
$(b_F^4\,a_0^2)^{1/5}$ only, which in Table~\ref{tab:kernels} is given as $E_F$.

As an example, consider the estimation of the forces in a Plummer sphere using
the F$_1$ softening kernel (see Table~\ref{tab:kernels}). Using the above
relations and the (analytic) values of $A$ and $B$ for a Plummer sphere with
unit mass and scale radius, we find ($G\equiv1$)
\[
        \moptF\approx 0.58\times N^{-4/5}.
\]
\subsubsection{Plummer softening}\label{sec:opt:plummer}
For Plummer softening, $a_0$ \eqi{ak} does not exist, because the softening
kernel is not compact, i.e.\ in the sense of \eqns{biases} the softening volume
(as measured by $a_0$) is infinite. As a consequence, the bias increases faster
than $\epsilon^2$. For centrally concentrated systems, I found from numerical
experiments that\footnote{\label{fn:plum}
        The reader may compare this with Fig.~4 of Athanassoula et al.\ (1998),
        which clearly gives a slope very near to 3, even though the authors of
        that paper use ${\rm bias}^2\propto\epsilon^4$ as asymptotic relation.}
$\isbF\propto\epsilon^{3.3}$ at $\epsilon\sim0.01$ scale radii, both for a
Plummer sphere and a Hernquist model as target. Using this empirical relation,
we get $\moptF\propto N^{-0.77}$ in good agreement with the numerical results of
Athanassoula et al.\ (2000), who find for $10^3\le N\le3\times10^5$ and a
Plummer sphere as underlying model:
\[
        \moptF\approx 0.45\times N^{-0.73}.
\]
This relation can be directly compared to that given above for spline softening.
Evidently, with ever larger $N$ Plummer softening becomes ever less effective
than compact kernels in reducing the \miseF\ (for small but historically
relevant values of $N$, the smaller coefficient seems to give Plummer softening
an edge).
\subsubsection{Kernels with $a_0=0$}\label{sec:opt:improved}
As we have seen in \eqn{bias:force}, the bias in the estimated force is, to
lowest order, equal to $-G\,\epsilon^2a_0\bnabla\!\rho$. One may actually
estimate the density in order to subtract this bias. This procedure, however, is
completely equivalent to using in the first place a kernel $\phi(r)$ for which
$a_0=0$. From \eqn{as:rho}, it follows that such a kernel corresponds to a
density $\eta(r)$ that must be negative at some $r$. For such a kernel,
\begin{equation}
        \miseF \approx A^\prime\,a_2^2\,\epsilon^8 
                + B\,b_F\,N^{-1}\epsilon^{-1}
\end{equation}
with
\begin{equation}
        A^\prime = \textstyle G^2\,M^{-1} \int\dx\;\rho(\bx)
                        \big[\bnabla\!\Delta\rho(\bx)\big]^2
\end{equation}
and $B$ given in \eqn{B} above. The softening that minimizes \miseF\ yields
\begin{subequations} \label{force:opt:compen} \begin{eqnarray}
        \epsilon_{\rm opt} &\approx& 8^{-1/9}\,(B/A^\prime)^{1/9}\,
                        (b_F/a_2^2)^{1/9}\,N^{-1/9},
\\[1ex] \label{force:opt:compen:mise}
        \moptF &\approx& 8^{-8/9}9\,(A^\prime\,B^8)^{1/9}\,
                        (b_F^8a_2^2)^{1/9}\,N^{-8/9}.
\end{eqnarray} \end{subequations}
This time, the dependence of the optimal \miseF\ on the kernel can be boiled
down to the constant $(b_F^8a_2^2)^{1/9}$, which is given in
Table~\ref{tab:kernels} as $E_F$. For the Plummer sphere as underlying model,
the kernel K$_1$ (see Table~\ref{tab:kernels}), \eqn{force:opt:compen:mise}
gives for large $N$ (in the same units as above)
\[
        \moptF \approx 1.00\times N^{-8/9}.
\]
\subsection{The choice of the kernel}\label{sec:kernel:opt}
After these considerations, we can answer the important question of the optimal
kernel $\phi(r)$. Below I list criteria, based on the above consideration and
decreasing in importance, which should be satisfied by the kernel $\phi$.
\newcounter{lll}
\setcounter{lll}{0}
\begin{list}{\arabic{lll}.}{\usecounter{lll} \leftmargin4mm 
\itemsep0pt plus 1pt
\parsep0pt plus 1pt
\topsep2pt plus 1pt
\labelwidth3mm \labelsep1mm \itemsep2ex}
\item   In order to reduce artificial small-scale noise, the estimated force must
	be everywhere continuous, i.e.\ the kernel must have a harmonic
	core\footnote{
		Actually, a super-harmonic core with $\phi^\prime\propto
		r^n$(for $r\ll\epsilon$) with integer $n>1$ is also acceptable,
		but gives a larger bias.},
        i.e.\ $\phi^\prime\propto r$ for $r\ll\epsilon$.
\item	In order to retain Newtonian gravity as close as possible with the
	above restriction, the force bias should be as small as possible. With
	the above results, this implies that the kernel should be compact, or,
	even better, have $\phi=r^{-1}$ (equivalent to $\eta=0$) for $r>r_0$,
	where $r_0$ is of the order of one.
\item   The kernel should not only yield continuous forces everywhere, but also
        continuous first, and possibly second, force derivative. This is 
        to facilitate accurate integration of the equations of motion.
\item   The kernel and its derivative should be easily evaluable.
\item   The kernel should result in the smallest possible mean integrated square
	forece error, \miseF.
\end{list}
Here, the exact order of the last three points is debatable. For instance, a
kernel that results in higher-order continuity at the price of being much more
expensive to evaluate and/or of resulting in much worse a force error is not
recommendable. Regarding point 4, it might be interesting to note that for
kernels that are polynomials in $r^2$, all mutual forces can be computed in
${\cal O}(N)$ operations (provided all mutual interactions are softened).
\subsubsection{Kernels with non-negative density}\label{sec:opt:cla:ass}
It is instructive to check the most widely used kernels against this list. All
of them satisfy the most important criterion 1, but the very popular Plummer
softening already fails at point 2: it is non-compact and produces a large bias,
i.e.\ gravitity is modified not only at small distances, but on all scales (see
Section~\ref{sec:opt:plummer}), which also results in large a \miseF\ (point 5).
The homogeneous-sphere (=Ferrers sphere of index $n=0$) softening proposed by
Pfenniger \& Friedli (1993) has only continuous forces, but discontinuous force
derivative and thus fails at point 3 (but satisfies 4), while the popular spline
softening introduced by Monaghan \& Lattanzio (1985) in the context of smoothed
particle hydrodynamics and advocated by Hernquist \& Katz (1989) also for force
estimation fails in the least important points 4 and 5 (the latter follows from
its efficiency $E_F$ being somewhat larger than for other compact kernels, see
Table~\ref{tab:kernels})\footnote{\label{note:spline}
	The rather complex functional form of the potential for the spline
	softening kernel originates from the necessity, in hydro- but not
	stellar dynamics, of an everywhere continuous second derivative of the
	density. An alternative kernel with this property is the $n=3$ Ferrers
	sphere, F$_3$ in Table~\ref{tab:kernels}, see also
	Appendix~\ref{app:kernels}.}.

Thus none of the popular softening kernels satisfies all criteria!
Table~\ref{tab:kernels} list these three kernels together with further possible
kernels. The Ferrers (1877) sphere kernels of integer index $n>0$ may be derived
(see Appendix~\ref{app:kernels}) as the simplest kernels (in the sense of being
low-order polynomials in $r^2$) that satisfy the first four of the above
criteria and are continuous in the $n$th force derivative.

\begin{figure}
	\ifpreprint
        \centerline{	\epsfxsize=70mm
			\epsfbox[28 158 444 717]{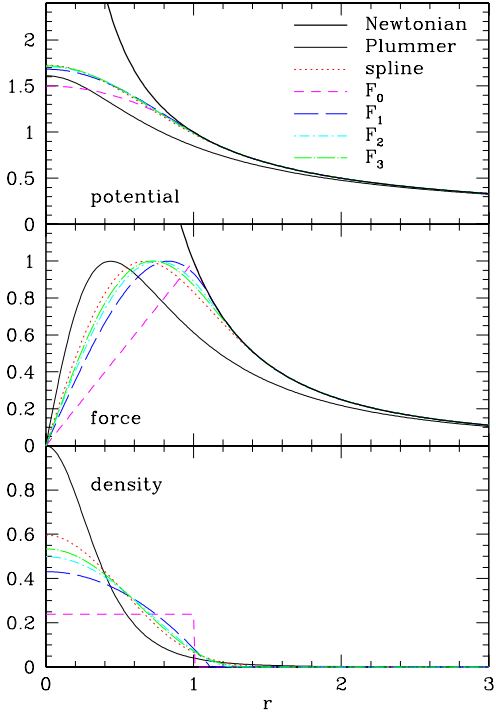}}
	\else \vspace*{2cm}\fi
        \caption[]{Potential, force, and density for Plummer and spline
		   softening kernels as well as the Ferrers (1877) sphere
	 	   kernels F$_n$. The softening lengths $\epsilon$ are scaled such
		   that the maximal force equals unity.
                \label{fig:kernels}}
\end{figure}

Figure~\ref{fig:kernels} plots the potential ({\em top}), force ({\em middle}),
and density ({\em bottom}) of these kernels and of Newtonian gravity. The
softening lengths $\epsilon$ are scaled such that the maximial force, i.e.\ the
artificial force scale introduced by softening (Newtonian gravity is scale
free), is the same for all kernels. Obviously, Plummer softening gives by far
the worst force approximation to Newtonian gravity (this holds for other ways of
scaling $\epsilon$, too). The F$_0$ (=homogeneous sphere) softening gives the
best force approximation, but at the price of an ugly discontinuity at
$r=\epsilon$.
The next best approximation to Newtonian gravity is the F$_1$ kernel (in the
field of density estimation also known as ``Epanechnikov'' kernel). If its
discontinuous second force derivative is a problem, one should use the F$_2$
kernel, while the spline kernel does not satisfy point 4 of the above list and
also compares somewhat worse to Newtonian gravity than does the F$_2$ kernel
(but see footnote \ref{note:spline}).
\begin{figure}
	\ifpreprint
        \centerline{	\epsfxsize=70mm
			\epsfbox[28 158 444 717]{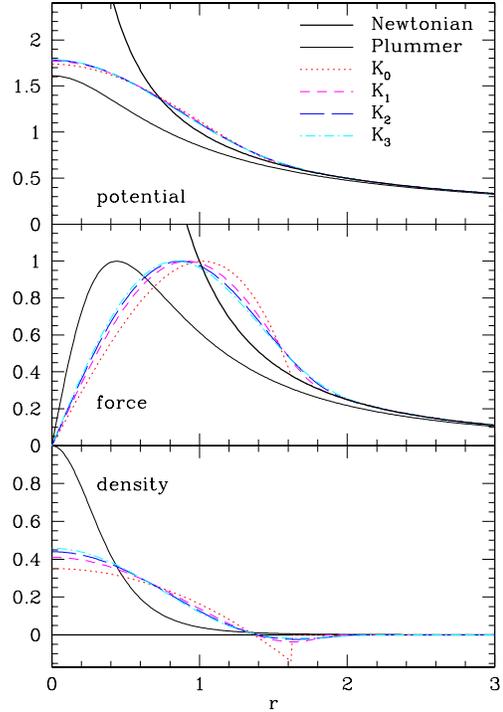}}
	\else \vspace*{2cm}\fi
        \caption[]{Like Figure~\ref{fig:kernels}, but for the kernels K$_n$ and
                   the Plummer kernel. Near their outer edge, these first
                   kernels have negative density and forces that exceed those of
                   Newtonian gravity. As a result, the under-estimation of the
                   force field due to too small forces near the origin is largely
                   compensated. \label{fig:kernels:impr}}
\end{figure} \par
One may want to derive the kernel that minizes the asymptotic force error,
i.e. $(b_F^2 a_0)$, subject to positivity and continuity. However, such an
optimal kernel is likely to be only marginally better than, say, the F$_1$
kernel (which is optimal for one-dimensional density estimation, see Silverman
1986), at the price of having a significantly more complex functional form.
\subsubsection{Kernels with bias compensation}\label{sec:opt:imp:ass}
In Appendix~\ref{app:kernels}, some simple kernels are derived which are
low-order polynomials in $r^2$, yield $a_0=0$, and satisfy all points in the
above list. These kernels, listed in Table~\ref{tab:kernels} as K$_n$, have
forces with continuous $n$th derivative. Figure~\ref{fig:kernels:impr} plots
their potential, force, and density in comparison with Newtonian gravity and the
Plummer kernel. While in the inner parts these kernels are harmonic, i.e.\ have
vanishing forces like the kernels shown in Fig.~\ref{fig:kernels}, their forces
exceed Newtonian gravity in the outer parts where their density is
negative. This largly compensates the under-estimation of gravity in the inner
parts and thus reduces the bias. As we shall see, these kernels prove to be
superior for the purpose of force estimation, while for density estimation they
are less useful because of their $\eta(r)$ not being non-negative.
\section{adaptive softening lengths and\\ \ifpreprint\hskip1.55em \fi
	 individual weights} \label{sec:ind}
To further improve the force estimation one may use locally adapted softening
lengths and/or individual weights. Both techniques aim at enhancing the
resolution in regions of high phase-space density by reducing $\epsilon$ or
increasing the number density of bodies in such regions. Here, I will formally
assess the effect of these possibilities on the force estimation.
\subsection{Definitions} \label{sec:ind:def}
\subsubsection{Individual weighting} \label{sec:ind:def:w}
The formalism behind this is to draw the initial positions $\{\bx_i,\bv_i\}$ not
from the DF $f(\bx,\bv)$ modelled, but from a {\em sampling\/} DF $f_{\rm s}$,
which is normalized to the same total mass $M$ as $f$. In this case, the bodies
have individual weights
\begin{equation} \label{mi}
        m_i = M\,N^{-1} \mu_i, \quad\mu_i\equiv\mu(\bx_i,\bv_i)
\end{equation}
with the weight function
\begin{equation} \label{mu}
	\mu(\bx,\bv) \equiv f(\bx,\bv)/f_{\rm s}(\bx,\bv),
\end{equation}
i.e.\ lighter bodies are used to compensate for oversampling, corresponding to
$f_{\rm s}$ being larger than $f$, and vice versa.

A potential problem with this method is an enhanced artificial heating of the
population of low-mass bodies, e.g.\ modelling a stellar disk, by two-body
interactions with high-mass bodies, e.g.\ modelling a dark halo. In order to
overcome this problem, I propose below to use larger individual softening
lengths for more massive particles.
\subsubsection{Adaptive softening}
	\label{sec:ind:def:eps}
For density estimation (see Silverman 1986), the usual procedure is to replace
$\epsilon$ in \eqns{kern} by $\epsilon\lambda_i$, resulting in the {\em adaptive
kernel estimators} (with individual weights $m_i$):
\begin{subequations} \label{adap} \begin{eqnarray} \label{adap:pot}
        \hPhi(\bx) &=& -{G\over\epsilon}\sum_{i=1}^N {m_i\over\lambda_i}\;
                \phi\left[{|\bx-\bx_i|\over\epsilon\lambda_i}\right],
\\ \label{adap:for}
        \hbF(\bx)  &=& {G\over\epsilon^2}\sum_{i=1}^N {m_i\over\lambda_i^2}\;
                \phi^\prime\left[{|\bx-\bx_i|\over\epsilon\lambda_i}\right]\;
                {\bx-\bx_i\over|\bx-\bx_i|}.
\end{eqnarray} \end{subequations}
So, $\epsilon$ no longer is a softening length, but the {\em softening
parameter\/} that controls the overall degree of softening. The $\lambda_i$ are
individual {\em bandwidths}, which are usually set to (Silverman 1986, p.~101)
$\lambda_i\propto \hrho_i^{-\alpha}$ where $\hrho_i$ is an estimate for the
density at $\bx_i$. The {\em sensitivity parameter\/} $\alpha$, a number
satisfying $0\le\alpha\le1$, determines how sensitive the local softening length
is to variations in the density.

In the presence of individual weights, one may better use
\begin{equation} \label{lambda}
	\lambda_i = \mu_i^{1/2}\,\left(\hn_i / \overline{n}\right)^{-\alpha},
\end{equation}
where $\hn_i$ are estimates of the local {\em number density\/} of bodies, while
$\overline{n}$ is a normalization constant, hereafter the arithmetic mean of the
$\hn_i$. With this formula for the bandwidth, the maximal force in the
neighbourhood of any body, $\bF_{\rm max}\propto m_i\lambda_i^{-2}$, is
independent of $m_i$. For $\alpha=0$ and $\mu_i\equiv1$ one obtains the standard
fixed-kernel estimators \eqb{kern}.

The number of bodies within a softening volume is approximately
\begin{equation}
        N_{\rm soft} \approx {4\pi\over3}\,\epsilon^3\lambda_i^3 n(x)
		     \approx {4\pi\over3}\,\epsilon^3\;\overline{n}^{3\alpha}
			\mu_i^{3/2}\;\hn(\bx)^{1-3\alpha}.
\end{equation}
Thus, the choice $\alpha=1/3$ results in $\mu_i^{-3/2}N_{\rm soft}$ being
approximately constant over the entire system. In this case, one may,
alternatively to \eqn{lambda}, determine the local softening length such that
the number of bodies within the softening volume is {\em exactly\/} proportional
to $\mu_i^{3/2}$ (or constant if the bodies are equally weighted). In this case,
the (fixed) number $\mu_i^{-3/2}N_{\rm soft}$ is the softening parameter.

In order to obtain the estimates $\hn_i$ for the number densities, one may use
the estimator
\begin{equation}
        \hn(\bx) = {1\over\epsilon^3} \sum_{i=1}^N {1\over\lambda_i^3}\;
                \eta\left[{|\bx-\bx_i|\over\epsilon\lambda_i}\right].
\end{equation}
with the bandwidths $\lambda_i$ from the previous time step\footnote{
	However, this method introduces a time asymmetry, which in turn can lead
	to unintended consequences. Therefore, one may use a simpler, and hence
	faster computable, estimate for the number density -- it is actually not
	important that the estimates $\hn_i$ usd in \eqn{lambda} are very
	accurate (see Silverman 1986). In a tree-code, for instance, one may use
	the mean number density within cells containing a predefined number of
	bodies.}.

If the method of individual weights is applied with $\alpha=0$, \eqn{lambda}
implies that the bodies have individual, though non-adaptive, bandwidths. Note
that this is by no means the only way incorporate $m_i$ into $\lambda_i$ --
other ways may work as well, but here I only consider \eqn{lambda}.

Hernquist \& Katz (1991) dubbed this sort of adaptive softening the `scattering'
method. An alternative is the `gathering' method, in which the softening length
depends not on the gravity source (and its position $\bx_i$), but on the sink.
A considerable disadvantage of both the scattering and the gathering method is
the implied violation of momentum conservation. However, momentum conservation
can easily be resurrected by using the same softening length $\epsilon_{ij}$ for
the force of body $i$ on body $j$ and vice versa. Traditional choices are
$\epsilon_{ij}^2=\epsilon^2(\lambda_i^2 +\lambda_j^2)$ (particularly useful in
case of Plummer softening), $\epsilon_{ij}=\epsilon\,(\lambda_i+\lambda_j) /2$
\cite{hb90}, or $\epsilon_{ij} = \epsilon\;{\rm min}(\lambda_i,\lambda_j)$. In a
simulation of a galaxy group with each galaxy's stellar component
represented by a single spheroidal particle, which was considerably
more massive than the halo bodies and therefore had larger bandwidth,
Barnes (1985) used $\epsilon_{ij}=\epsilon\;{\rm max}(\lambda_i,
\lambda_j)$, to avoid strong heating by individual halo body interactions.
\subsection{Errors} \label{sec:ind:errors}
\subsubsection{The bias} \label{sec:ind:bias}
The bias of the potential estimated using adaptive softening and individual
weights is (see Appendices \ref{app:adap} and \ref{app:ind})
\begin{equation} \label{ind:bias:pot}
        {\rm bias}_\bx(\hPhi) = a_0\,G \epsilon^2\rho(\bx)\,\overline{\mu}(\bx)\,
        \left[{n(\bx)\over\overline{n}}\right]^{-2\alpha} \;+\;{\cal O}(\epsilon^4),
\end{equation}
where $\overline{\mu}\equiv\rho^{-1} \int\dv f \mu$ is the mass-weighted average
of $\mu$ of all trajectories passing through $\bx$. This result is independent
of the method used to assign $\epsilon_{ij}$ (it affects only the higher-order
terms), while ${\rm bias}_\bx(\hbF)=\bnabla{\rm bias}_\bx(\hPhi)$.

Let us first consider $\mu\equiv1$ (equal weighting). In this case,
$n\propto\rho$ and therefore ${\rm bias}_\bx(\hbF)\propto a_0 \epsilon^2
(1-2\alpha)\rho^{-2\alpha}\bnabla\rho$. Compared to \eqn{bias:force}, the bias
is reduced everywhere by the factor $(1-2\alpha)$ (for $\alpha=1/2$ we even have
formally ${\rm bias}(\hbF)= {\cal O}(\epsilon^4)$, but in this case the
derivation of \eqn{ind:bias:pot} becomes inconsistent, see
Appendix~\ref{app:adap}). Moreover, the factor $[\rho(\bx)/\overline{\rho}
]^{-2\alpha}$ decreases the bias in high-density regions and increases it in
low-density regions, such that the density-weighted \isbF\ is reduced.

If in addition $\mu\not\equiv1$, the bias is further reduced. First $n(\bx)$ is
a steeper function than $\rho(\bx)$ and hence $n^{-2\alpha}$ is shallower
resulting in a smaller gradient (=${\rm bias}_\bx(\hbF)$). Second, if the mean
weight $\overline\mu$ is smaller in high-density regions, this produces a
smaller gradient in the same way.

\subsubsection{The variance} \label{sec:ind:var}
For the variance of the force, we obtain (see Appendices \ref{app:adap} and
\ref{app:ind})
\begin{equation}
        N{\rm var}_\bx(\hbF) = b_F G^2 M\epsilon^{-1}
		\rho(\bx)\,\overline{\mu^{1/2}}(\bx)
		\left[{n(\bx)\over\overline{n}}\right]^\alpha 
	+ {\cal O}(\epsilon^0).
\end{equation}

Let us again first consider the case of equal weights. Then $n\propto\rho$, and
${\rm var}_\bx(\hbF)\propto\rho^{1+\alpha}$, i.e.\ in contrast to the bias, the
variance is enhanced in high density regions, because the softening length is
reduced there. The corresponding increase in \ivF\ is smaller than the decrease
in \isbF, because the density contrast $\rho/\overline{\rho}$ enters the
integral for the evaluation of \ivF\ only to the power $\alpha$, while in the
evaluation of \isbF\ it enters to the power of $-4\alpha$. However, since at the
optimal softening, $\ivF=4\,\isbF$, the decrease in the \isbF\ must be more than
four times the increase in the \ivF, in order to yield an overall improvement in
the optimal \miseF\ (which occurs at larger $\epsilon$ than for the
corresponding fixed-kernel estimate). Due to this effect, we do not necessarily
expect a dramatic improvement of the force estimation from adaptive rather than
fixed kernel estimators.

Consider next the situation for individual weights. We may replace $n$ by
$\rho\overline{\mu^{-1}}$ to find that, compared to the equal-weight case, the
variance is multiplied by $\overline{\mu^{1/2}\,}\overline{\mu^{-1}}^{\alpha}$.
For $\mu_i$ not too far from unity, $\overline{\mu^{a}}\sim\overline{\mu}^{a}$
and the factor becomes $\sim\overline{\mu}^{1/2-\alpha}$. That is, for
$\alpha\la1/2$, the variance of the force is {\em reduced} in regions of smaller
than average weights, decreasing the global measure \ivF.

Thus, when using individual weights in conjunction with adaptive softening, both
variance and bias of the force are reduced in high-density regions and therefore
also the mass-weighted averages \ivF\ and \isbF\ as well as the overall force
error \miseF.

\section{Numerical Experiments}\label{sec:exper}
So far, we have studied the properties of the softening purely analytically in
the asymptotic limit of small $\epsilon$ and large $N$. In order to study the
behaviour at finite $N$ and to assess the applicability of the asymptotic
relations, we will now turn to numerical experiments. To this end, we compute
the bias and the variance of the force at a hundred radii, $r_k$, equidistant in
the enclosed mass, i.e.\ $M(<r_k)=(k+{\textstyle{1\over2}})/100$, and estimate
the \isbF\ and \ivF\ as averages over these points.
\subsection{The targets}\label{sec:exper:targets}
The comparisons will be made for two spherically symmetric models. The first is
the Plummer (1911) sphere with density and potential
\begin{equation} \label{plummer:sphere}
        \rho(r)={3Mr_s^2\over4\pi(r_s^2+r^2)^{5/2}},\quad
        \Phi(r)=-{GM\over\sqrt{r_s^2+r^2}}.
\end{equation}
(hereafter $G\equiv M\equiv r_s\equiv 1$), which has the advantage that most
quantities in the above estimates can be evaluated analytically. The Plummer
sphere has a harmonic core with central frequency $\omega=1$, i.e.\ $t_{\rm
dyn}=2\pi$, in these units.  I will also use the Hernquist (1990) model
\begin{equation} \label{hernquist:model}
        \rho(r)={Mr_s\over2\pi r(r_s+r)^3},\quad
        \Phi(r)=-{GM\over r_s+r}.
\end{equation}
This model resembles the properties of galaxies better, since it has a central
density cusp, where some of the asymptotic estimates made in
Section~\ref{sec:errors} are not valid.

The Hernquist model has a central force $\bF\to-\bx/r$ as $r\equiv|\bx|\to0$.
Any softening yields a continuous force field and therefore inevitably gives
$\langle\hbF\rangle=0$ in the centre, resulting in a 100\% central force
bias. Since this error is restricted to the softening volume where the density
scales as $r^{-1}$, its contribution to the overall \isbF\ scales as
$\epsilon^2$. Thus, because the \isbF\ for a non-cusped system scales as
$\epsilon^4$, we expect the central bias to dominate the overall error for
sufficiently small $\epsilon$, i.e.\ large $N$.

This is a generic problem of any cusped stellar system: the force field of a
steep cusp cannot be resolved with any softening\footnote{
        For density cusps steeper than that of the Hernquist model, i.e.\
        $\rho\propto r^{-\gamma}$ as $r\to0$ with $\gamma>1$, which are observed
        in many early-type galaxies, the central force diverges resulting in a
        formally infinite force bias, unless $\epsilon=0$.}.
Moreover, a cusp creates further problems for \Nbd\ simulations, since the
dynamical time scale becomes arbitrary small as $r\to0$. This means that one
cannot even hope to properly model a cuspy stellar system with any \Nbd\ method,
but must accept a small artificial core with size of the order of the softening
length\footnote{ A steep power-law density cusp can be modelled down to
$r\sim0.1\epsilon$ by setting up initial velocities in equilibrium with the
softened potential \cite{barnes98}. However, this amounts to modifying the DF on
scales $r\la\epsilon$, not $r\la0.1\epsilon$. Thus, also in this approach the
\Nbd\ simulation does not match the stellar system modelled on a scale smaller
than $\sim\epsilon$.}.  Given this generic problem of all softening techniques,
it makes sense to compare their performance only at $r>\epsilon$, i.e.\ ignore
the problems at $r\to0$. This is done automatically by our estimate of the
\isbF, since the smallest radius for which we compute the bias is non-zero (it
contains 0.005 of the mass), i.e.\ larger than $\epsilon$ for small softening.

\subsection{Fixed kernel estimates}\label{sec:exper:fixed}
\subsubsection{Plummer sphere as target}\label{sec:exper:fixed:P}
\begin{figure}
        \ifpreprint
	\centerline{\epsfxsize=85mm \epsfbox[39 157 313 715]{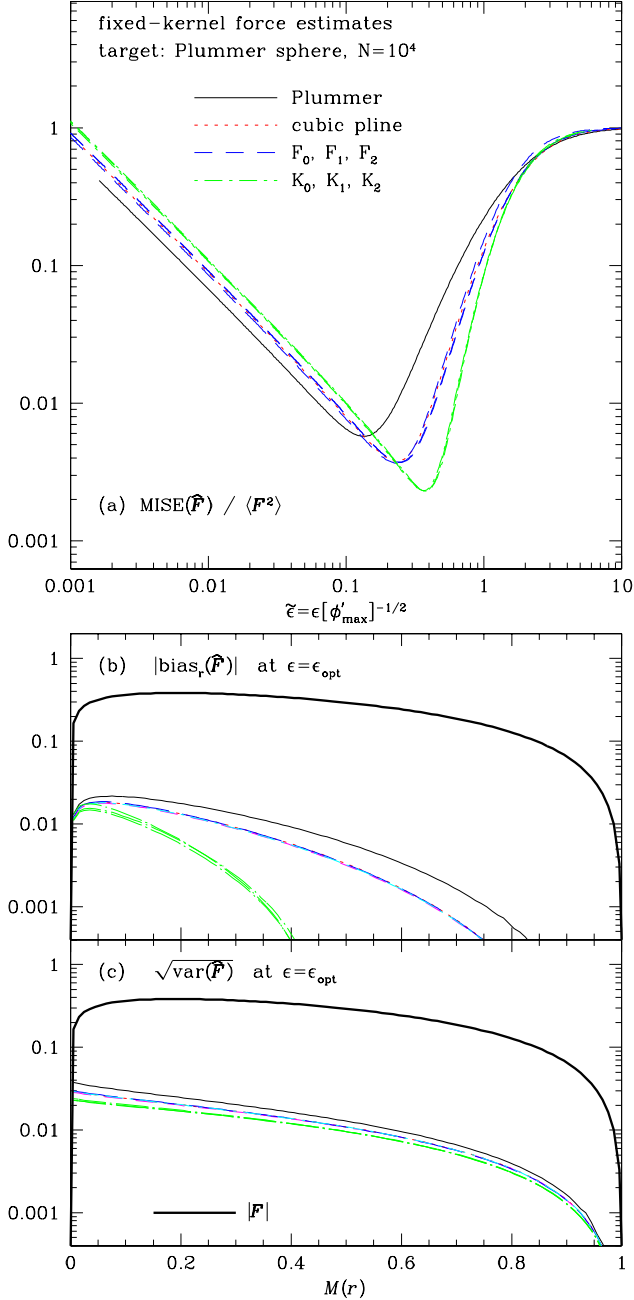}}
	\else \vspace*{2cm}\fi
        \caption[]{Fixed-kernel force estimate for most kernels of
        Table~\ref{tab:kernels} applied to $N=10^4$ positions drawn from a
        Plummer sphere. In the top panel (a), the mean integrated squared force
        error, \miseF, normalized by the averaged squared force of the Plummer
        sphere, is plotted vs.\ the scaled softening length $\tilde{\epsilon}$.
        The middle and bottom panels (b and c) give the radial run of the ${\rm
        bias}_\bx(\hbF)$ and the dispersion, $[{\rm var}_x(\hbF)]^{1/2}$, at the
        value of $\epsilon$ that minimizes the \miseF. For comparison, the bold
        curve represents the true force in a Plummer sphere.
        \label{fig:mise:P:fix}}
\end{figure}
For most softening kernels of Table~\ref{tab:kernels} (or
Figures~\ref{fig:kernels} and \ref{fig:kernels:impr}),
Figure~\ref{fig:mise:P:fix}a plots the \miseF, normalized by the averaged
squared force
\begin{equation}
        \langle\bF^2\rangle \equiv M^{-1} \int\dx\;\rho(\bx)\;\bF^2(\bx)
\end{equation}
of the Plummer sphere, for $N=10^4$ bodies versus the {\em scaled softening
length}
\begin{equation} \label{teps}
        \tilde{\epsilon} = \epsilon\,\big[\phi^\prime(r_{{\rm max}\,F})
		\big]^{-1/2}.
\end{equation}
The effect of this scaling is that the maximal force generated by a body of mass
$m$ is $m/\tilde{\epsilon}^2$, independent of the kernel (the distance from the
body at which this maximal force occurs is different for the different kernels,
see Figs.~\ref{fig:kernels} and \ref{fig:kernels:impr}).

In Figure~\ref{fig:mise:P:fix}a, the asymptotic behaviour at small and large
softening lengths is immediately apparent. At small $\epsilon$, the integrated
variance, \ivF, dominates the error budget, yielding a divergence as
$\epsilon^{-1}$ for $\epsilon\to0$.  At fixed $\tilde{\epsilon}$, Plummer
softening yields the smallest \ivF, which is related to its long-range
softening.

At large $\epsilon$, the \miseF\ is dominated by the integrated squared bias,
\isbF, which increases steeply until $\tilde{\epsilon}\simeq1$, beyond which it
saturates at $\miseF=\langle\bF^2\rangle$, corresponding to a vanishing
estimated force. As expected from our asymptotic results, the \isbF\ grows like
$\epsilon^4$ for the compact kernels with non-negative density (cubic spline \&
F$_n$). For these kernels, the optimal scaled softening lengths and the
resulting minimal \miseF\ are very similar. The latter is a consequence of the
fact that the combination $E_F=b_F^4a_0^2$, through which the \moptF\ depends on
the kernel, does not vary much between those kernels (see
Table~\ref{tab:kernels}).

For Plummer softening, the \isbF\ is much larger (up to one order of magnitude)
and grows approximately like $\epsilon^{3.3}$. As a result, the optimal
softening length, $\tilde{\epsilon}_{\rm opt}$, is shorter (and hence the
maximal force generated by any body larger) and the ${\rm IV}_{\rm opt}(\hbF)$
as well as \moptF\ are larger than for compact kernels.

For the kernels K$_n$, which have partly-negative $\eta(r)$ and $a_0=0$, the
\isbF\ is smaller than for the other kernels and grows like $\epsilon^8$ (for
sufficiently small $\epsilon$). The \ivF\ is only slightly larger than for the
non-negative kernels, so that the minimal force error, \moptF, is significantly
smaller than for other kernels and occurs at larger $\tilde{\epsilon}$, i.e.\
the maximum force generated by a single body at optimal softening is smallest
for these kernels.

Figures~\ref{fig:mise:P:fix}b and \ref{fig:mise:P:fix}c show the radial run of
the bias and dispersion, $[{\rm var}_\bx(\hbF)]^{1/2}$, for the optimal
softening, corresponding to the minima in Figure~\ref{fig:mise:P:fix}a. The
lines for the four compact kernels with non-negative $\eta(r)$ almost overlay
each other, as do the lines for the kernels K$_n$. This is exactly, what is
expected from the asymptotic relations of Section~\ref{sec:errors}, because of
the similarity of the factors $E_F=b_F^4a_0^2$ and $b_F^8a_2^2$,
respectively. In all cases, the dispersion is larger than the bias, in agreement
with our expectations, based on the asymptotic relations, that $\ivF=4\isbF$
(for compact kernels with $a_0\neq0$). While in the inner parts of the Plummer
sphere the bias contributes still significantly to the total error, in the outer
parts the bias is neglible compared to the variance.  This is true especially
for the kernels K$_n$, which were designed to reduce the bias.
\subsubsection{Hernquist model as target}\label{sec:exper:fixed:H}
\begin{figure}
        \ifpreprint
        \centerline{\epsfxsize=85mm \epsfbox[39 157 313 715]{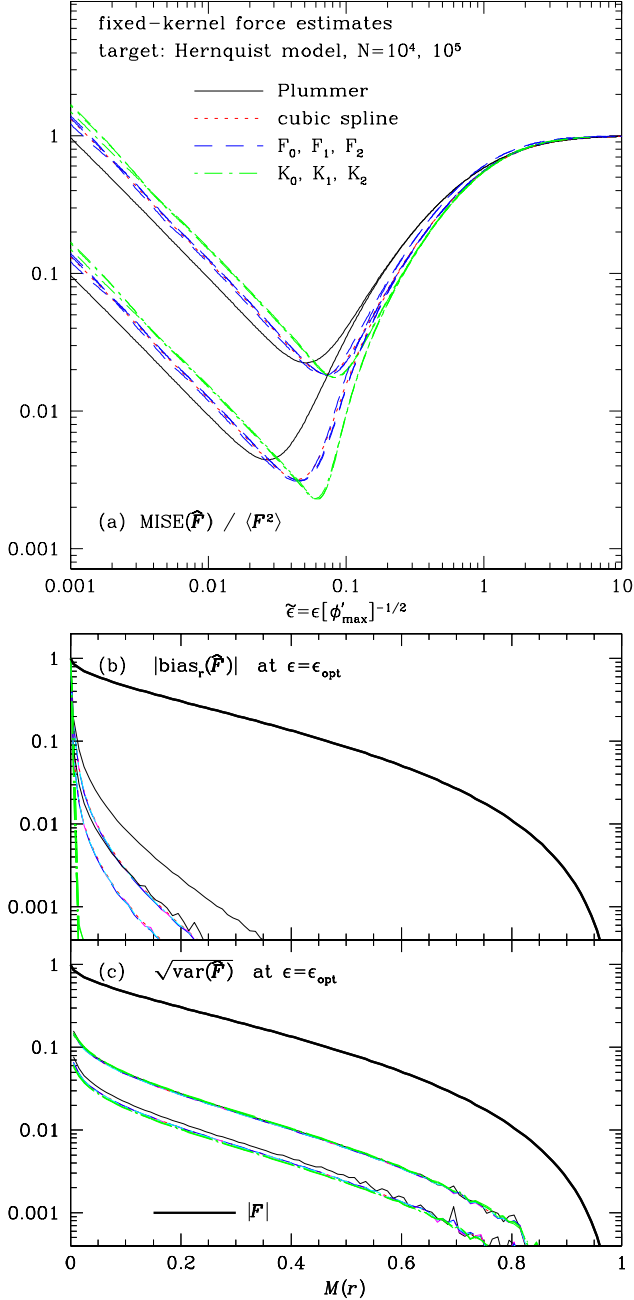}}
	\else \vspace*{2cm}\fi
        \caption[]{As Fig.~\ref{fig:mise:P:fix} but for the Hernquist model as
        target with $N=10^4$ (upper curves) and $N=10^5$ (lower curves).
        \label{fig:mise:H:fix}}
\end{figure}
For $N=10^4$ and $10^5$, Figure~\ref{fig:mise:H:fix} shows the results of
fixed-kernel force estimation applied to a Hernquist model. Compared with the
situation for the Plummer sphere (Fig.~\ref{fig:mise:P:fix}), the \ivF\ (i.e.\
the MISE at small $\tilde{\epsilon}$) is only marginally larger at any
$\tilde{\epsilon}$, as expected from the asymptotic relation: after
normalisation by averaged squared force, the coefficient $B$ in \eqn{B} is 20\%
larger than for the Plummer sphere. The \isbF\, on the other hand is
significantly larger than for the Plummer sphere. This has the consequence that
also the minimal \miseF\ is larger and occurs at smaller $\tilde{\epsilon}$ than
for the Plummer sphere. In particular, at $N=10^4$, the relative rms force error
$(\moptF/\langle\hbF\rangle)^{1/2}$ is as large as 14\%, and still about 5\% at
$N=10^5$.

The differences in the \moptF\ between the various kernels are small at $N=10^4$
but become clearer at $N=10^5$: the improvement in increasing $N$ by an order of
magnitude is significantly larger for the compact kernels, in particular K$_n$,
than it is for Plummer softening.

As already discussed in Section~\ref{sec:exper:targets}, the central cusp of the
Hernquist model causes a severe difficulty for any \Nbd\ method. The innermost
radius for which we compute the bias and variance is $\approx0.08$, and we thus
expect the estimated \isbF\ to scale as $\epsilon^2$ for $\epsilon$ larger than
this. The corresponding change in the slope is evident in
Fig.~\ref{fig:mise:H:fix}a.

The radial runs of ${\rm bias}_\bx(\hbF)$ and $[{\rm var}_\bx(\hbF)]^{1/2}$ in
Figs.~\ref{fig:mise:H:fix}b and c, respectively, are most interesting. In
contrast to the situation for the Plummer sphere, the variance dominates at all
body positions, except for the innermost few percent, especially for the kernels
K$_n$\footnote{
        This is related to a strange property of the Hernquist model:
        $\bnabla\Delta\rho$, which in the asymptotic limit is related to the
        force bias for these kernels (see\ \eqn{bias:force}), contains a part
        $\propto\bnabla\delta(\bx)$. For finite softening length this is tapered
        to a function with some finite width of the order of $\epsilon$.},
where the cusp results in a large bias.
\subsection{Adaptive kernel estimates}\label{sec:exper:adaptive}
Figures~\ref{fig:mise:H:fix}b and \ref{fig:mise:H:fix}c nicely show the dilemma
of fixed-kernels estimates: the large central ${\rm bias}_\bx(\hbF)$ requires,
in order to achieve the {\em global\/} balance between \isbF\ and \ivF, a small
softening length, which in turn results in a large ${\rm var}_\bx(\hbF)$ in the
outer parts. It would be much better to balance bias and variance {\em
locally\/}, in order to obtain an optimal mean square force error at every
position $\bx$. By choosing the local softening length in proportion to
$\rho^{-\alpha}$, the adaptive kernel estimates do not quite reach this goal,
but certainly go a good way in the right direction.

There are many parameters which one may vary in adaptive kernel estimates: the
kernels for the force and density estimators and their respective softening
lengths as well as the sensitivity parameter $\alpha$. However, as we have seen
in the fixed kernel estimates, the Plummer kernel is clearly inferior to the
other kernels discussed, and there is not much difference between the various
positive definite compact kernels or between the kernels K$_n$. Therefore, I
will restrict the experiments to the F$_1$ and K$_1$ kernels for the force
estimation, while the density is always estimated by the F$_1$ kernel. Since the
variance of the estimated density \eqi{var:density} increases much faster at
small $\epsilon$ than that of the estimated force, the optimal softening
parameter for density estimation is always larger than that for force
estimation. Therefore, I use twice the softening scale for the evaluation of
$\hrho(\bx_i)$ than in the force estimator.
\begin{figure}
        \ifpreprint
        \centerline{	\epsfxsize=85mm
			\epsfbox[39 157 313 715]{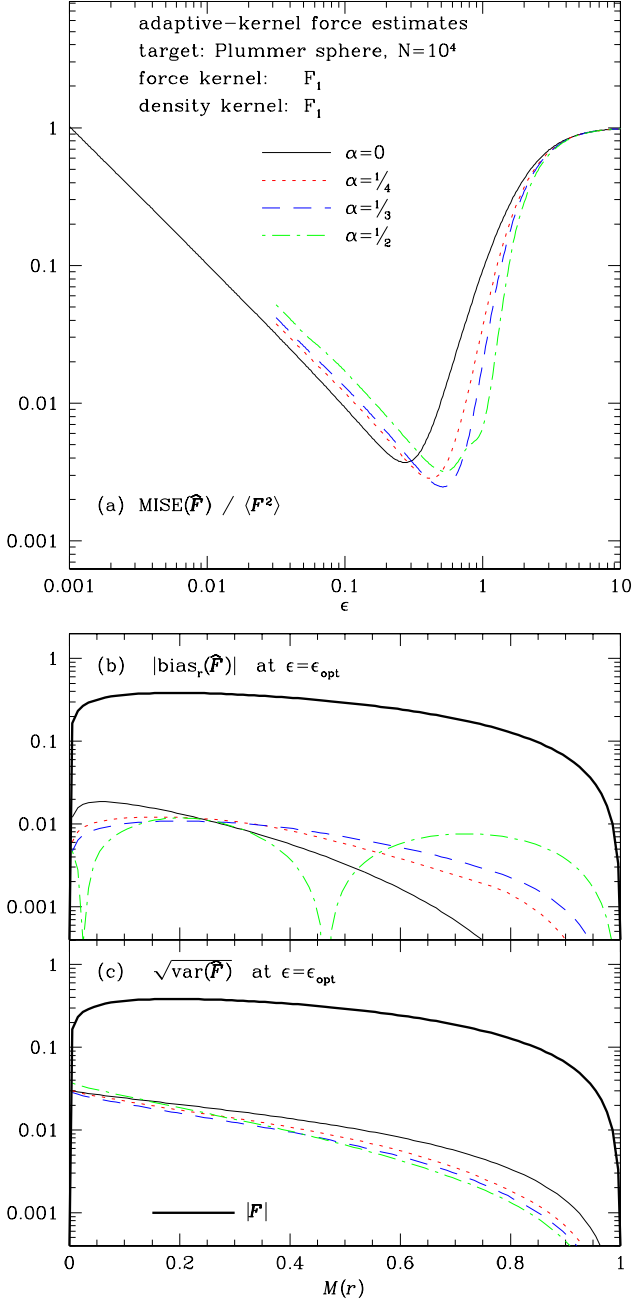}}
	\else \vspace*{2cm}\fi
        \caption[]{As Figure~\ref{fig:mise:P:fix} but for an adaptive-kernel
        estimator based on the F$_1$ kernel. The softening scale of the density
        estimator has been set to twice that of the force estimator. Note that a
        sensitivity parameter of $\alpha=0$ corresponds to the fixed-kernel
        estimator of Fig.~\ref{fig:mise:P:fix}. \label{fig:mise:P:ada:ep}}
\end{figure}
\begin{figure}
        \ifpreprint
        \centerline{	\epsfxsize=85mm
			\epsfbox[39 157 313 715]{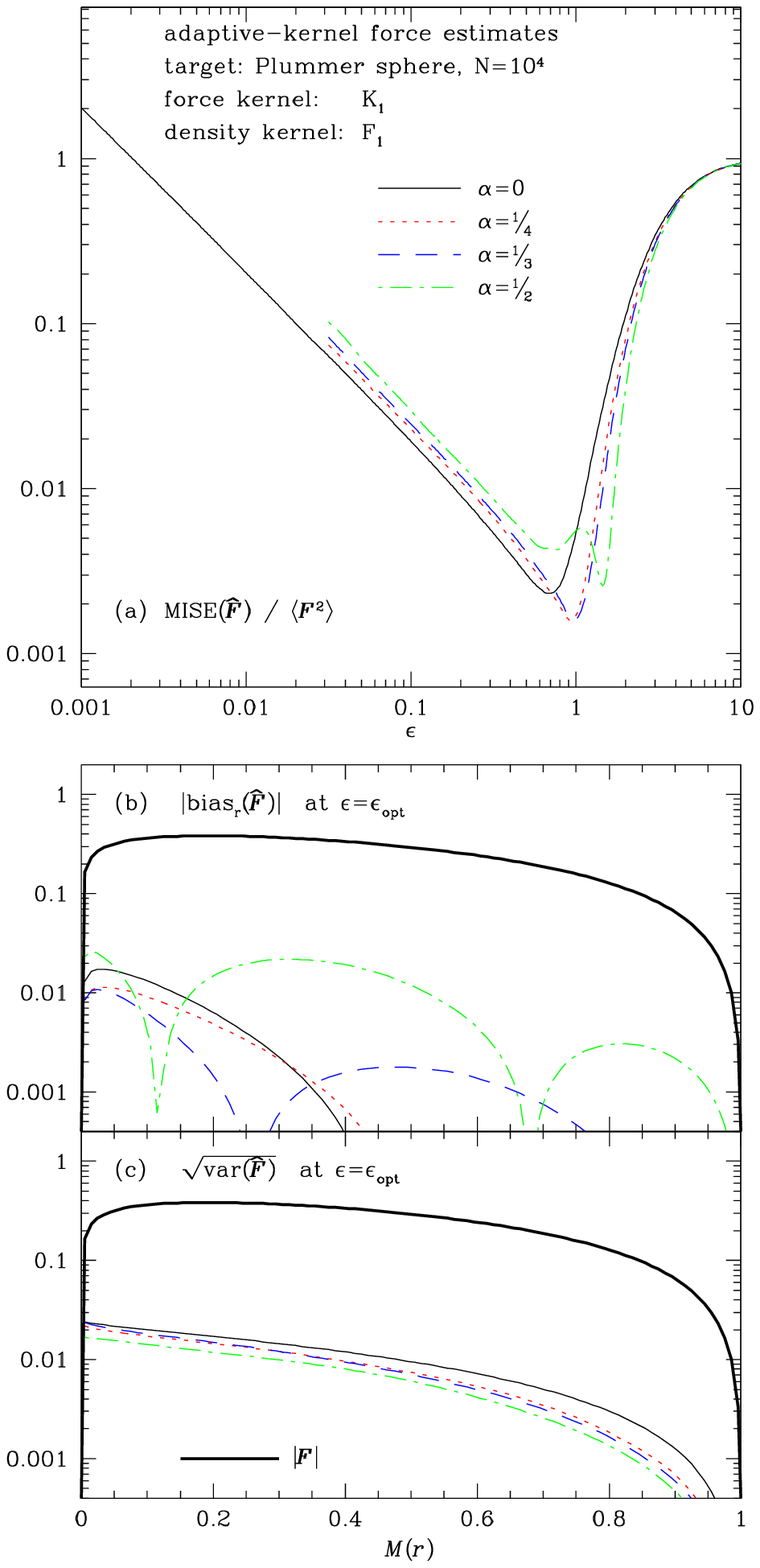}}
	\else \vspace*{2cm}\fi
        \caption[]{As Figure~\ref{fig:mise:P:ada:ep} but using the kernel K$_1$
        in the force estimator \label{fig:mise:P:ada:k}}
\end{figure}
\begin{figure}
        \ifpreprint
        \centerline{
			\epsfxsize=85mm
			\epsfbox[39 157 313 715]{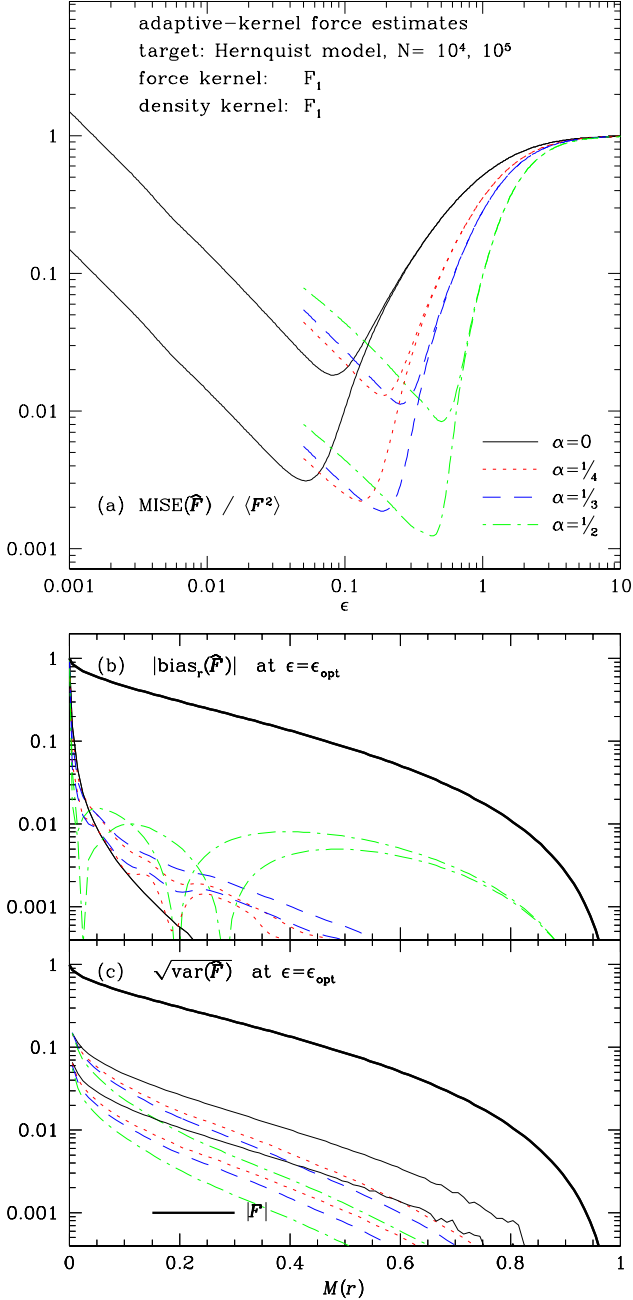}}
	\else \vspace*{2cm}\fi
        \caption[]{As Figure~\ref{fig:mise:H:fix} but for an adaptive-kernel
        estimator based on the F$_1$ kernel. The softening scale of the density
        estimator has been set to twice that of the force estimator.
        \label{fig:mise:H:ada:ep}}
\end{figure}
\begin{figure}
        \ifpreprint
        \centerline{
			\epsfxsize=85mm
			\epsfbox[39 157 313 715]{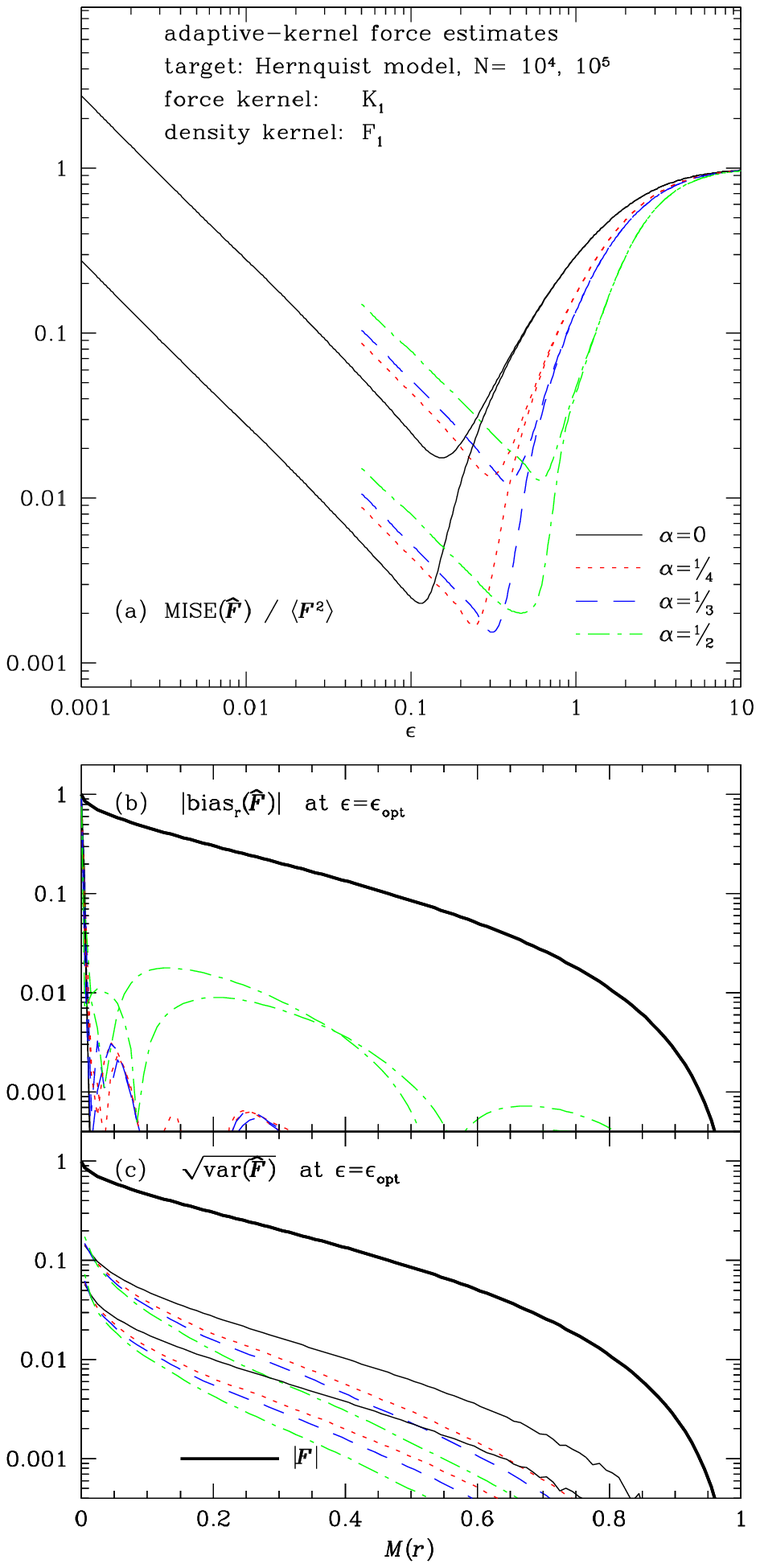}}
	\else \vspace*{2cm}\fi
        \caption[]{As Figure~\ref{fig:mise:H:ada:ep} but using the kernel K$_1$
        in the force estimator \label{fig:mise:H:ada:k}}
\end{figure}
\subsubsection{Plummer sphere as target}\label{sec:exper:adaptive:P}
For various choices of the sensitivity parameter $\alpha$ \eqi{lambda},
Figures~\ref{fig:mise:P:ada:ep}a and \ref{fig:mise:P:ada:k}a show the run with
$\epsilon$ of the \miseF\ obtained with the F$_1$ and K$_1$ kernels,
respectively, from $N=10^4$ positions drawn from a Plummer sphere. The curves
for $\alpha=0$ are identical to those of the corresponding kernel in
Fig.~\ref{fig:mise:P:fix}.  Adapting the local softening improves the accuracy
of the force estimation: at $\alpha=1/3$ the \moptF\ has decreased by a factor
of $\sim1.6$, because the bias is substantially reduced, while the variance has
been only slightly increased, in agreement with the expectations from theory in
Section~\ref{sec:ind:errors}.

Out of the numbers tested for $\alpha$, the value $1/3$ results in the best
estimate; the case $\alpha=1/2$ will be discussed seperately in
Section~\ref{sec:exper:adaptive:alpha}.

It is very instructive to consider the radial run of bias and variance at
$\epsilon=\epsilon_{\rm opt}$ in the lower panels of
Figures~\ref{fig:mise:P:ada:ep} and \ref{fig:mise:P:ada:k}. Compared to the
fixed-kernel estimates ($\alpha=0$), the adaptive-kernel estimates have reduced
variance, while the run of the bias is much shallower, exceeding that for
$\alpha=0$ at large radii and running below it at small radii. Together this
gives a better local balance between bias and variance and a local mean square
force error, ${\rm MSE}_\bx(\hbF)$ everywhere smaller than for the fixed-kernel
estimators.

It should be noted that, in case of the compensating kernel K$_1$, already
$\alpha=1/4$ results in a significant error reduction. For $\alpha=1/3$ and
$\alpha=1/2$ (here also for F$_1$), the bias changes sign. In these cases, the
bias is a rather complicated function of radius, the asymptotic form of which
has not been derived above.
\subsubsection{Hernquist model as target}\label{sec:exper:adaptive:H}
Figures~\ref{fig:mise:H:ada:ep}a and \ref{fig:mise:H:ada:k}a plot the \miseF\
versus $\epsilon$ when applying the adaptive-kernel force estimator with,
respectively, the F$_1$ and K$_1$ kernel to $N=10^4$ and $10^5$ positions drawn
from the Hernquist model. As anticipated from the asymptotic relations in
Section~\ref{sec:ind:errors}, the adaption reduces the \isbF\ (dominating the
\miseF at large $\epsilon$) substantially, much more than for the case of a
Plummer sphere as target. However, the increase of the \ivF\ is also quite
large, so that the \moptF\ is reduced by about the same amount as for the
Plummer target.  A check with Figs.~\ref{fig:mise:H:ada:ep}b\&c and
\ref{fig:mise:H:ada:k}b\&c shows that both \isbF\ and \ivF\ are completely
dominated by the large contributions from the innermost few per cent of the
mass.

Interestingly, there is a difference between the kernels F$_1$ (with $a_0\neq0$)
and K$_1$ (with $a_0=0$): for the latter the adaption does not reduce the
\moptF\ as much as for the F$_1$ kernel, and, as already found with the 
Plummer model as target (cf.\ Figure~\ref{fig:mise:P:ada:k}), the difference
between $\alpha=1/4$ and $1/3$ is marginal.
\subsubsection{What is the optimal $\alpha$?}\label{sec:exper:adaptive:alpha}
Special attention must be given to the case $\alpha=1/2$, for which the
asymptotic relation \eqb{ind:bias:pot} breaks down. The behaviour both of the
$\miseF$ and of the ${\rm bias}_\bx(\hbF)$ become strange at this choice of the
sensitivity parameter. For example, in Fig.~\ref{fig:mise:H:ada:ep},
$\alpha=1/2$ gives the smallest \moptF, but the bias reaches $\sim10\%$ in the
outer regions of the model, where it usually is negligible. In other cases,
$\alpha=1/2$ results in \miseF\ that behaves very odd as function of $\epsilon$,
sometimes even with two minima (see Fig.~\ref{fig:mise:P:ada:k}a), and in most
cases do not yield the smallest force error found.

The peculiar properties of $\alpha=1/2$ have been noticed in the context of
density estimation a number of times (Scott 1992), in particular, that for large
$N$ such estimates can be worse than fixed-kernel estimates -- in agreement with
Figure~\ref{fig:mise:P:ada:k}. This suggests that the behaviour of
adaptive-kernel estimates with $\alpha=1/2$ is not well understood (even in the
well-studied context of density estimation), and may thus be a topic for further
investigations.

In the face of these results, the choice $\alpha=1/2$ cannot be recommended.
However, a value of $1/3$ for the sensitivity parameter, corresponding to a
roughly constant number of bodies within the local smoothing volume, seems to
work well: it reduces the \moptF\ by a factor of about 1.6, the precise value
depending on the circumstances (kernel \& underlying model).
\section{Discussion}\label{sec:disc}
\subsection{Relevance for other than direct-summation codes}
\label{sec:disc:relev}
In our analysis, we have assumed that the forces are evaluated by direct
summation, which is a rather CPU time intensive technique. Consequently, cheaper
but approximate methods are favoured by most \Nbd\ engineers, with the exception
of GRAPE \cite{ebi93}, which reaches a higher performance by wiring
\eqn{plummer-soft} into the hardware.

The force error arising from these approximate methods, some of which combine
softening and approximation into one process, follows not necessarily the
relation \eqb{miseF:asym}. Nonetheless, the \miseF\ is a function of $N$ because
of ${\rm var}(\bF)\propto N^{-1}$. Of course, the force error also depends on
the parameters of the method (including the softening length $\epsilon$). Hence,
there exist optimal values for these parameters in the sense of minimizing the
\miseF\ at given $N$.

This implies the necessity to change the parameters of the method when
increasing $N$. However, this fact is often ignored by users of such techniques,
who keep the parameters constant when increasing $N$ and then claim their method
needs only ${\cal O}(N)$ or less operations for the force approximation. With
such a practice, the force error will eventually be dominated by the bias, and
increasing $N$ does not improve it at all. Moreover, the method becomes
statistically inconsistent in the sense that in the limit $N\to\infty$ the
estimated force does not converge to the true force. These last remarks apply in
particular to the expansion and grid-based techniques discussed below.
\subsubsection{The tree code}
An important method is the Barnes \& Hut (1986) tree code and its clones: the
force is evaluated by direct summation for nearby bodies only, whereas distant
bodies are grouped together and their gravity is computed via a low-order
multipole expansion. This method reduces the computational costs to ${\cal
O}(N\ln N)$ or even ${\cal O}(N)$ \cite{d00}.

The ignorance of the detailed structure of distant parts of the system
introduces some additional errors and one objective of the \Nbd\ experimenter
must be to keep these errors always smaller than those inherent to the force
estimation (by softening). The results of Athanassoula et~al.\ (1998) suggest
that this condition is indeed satisfied for common-practice applications, which
essentially implies that the results of this paper can be directly applied to
the tree code.
\subsubsection{Expansion techniques}
Another method is based on an expansion of density and potential in a set of
bi-orthogonal basis functions, ${\it\Phi}_{\bmath{n}}(\bx)$:
\begin{equation} \label{expand}
        \hPhi(\bx) = {\textstyle\sum}_{\bmath{n}} A_{\bmath{n}}
			{\it\Phi}_{\bmath{n}}(\bx),\quad
        A_{\bmath{n}}  = {\textstyle\sum_i} m_i\,{\it\Phi}_{\bmath{n}}(\bx_i).
\end{equation}
where the second equality follows from the bi-orthogonality relation
$4\pi\int\dx\,{\it\Phi}_{\bmath{m}}\bnabla^2{\it\Phi}_{\bmath{n}}=
\delta_{\bmath{n}\bmath{m}}$. The sum over $\bmath{n}$ in \eqn{expand} includes
terms up to some maximum $\bmath{n}_{\rm max}$ and contains ${\cal O}(n_{\rm
max}^3)$ terms in total.

This technique is important for special situations, in which the stellar system
is closely matched, at all times, by the low-order basis functions. For example,
when studying the stability of equilibrium systems of axial or polar symmetry,
this is the method of choice \cite{es95}. For less specialized investigations,
the stellar system is not always well matched by the low-order basis functions,
and the technique is essentially useless, because in this case $n_{\rm max}
\propto N^{1/3}$ for optimal force estimation, resulting in computational costs
that are ${\cal O}(n_{\rm max}^3\,N)={\cal O}(N^2)$.

To see that $n_{\rm max} \propto N^{1/3}$ for the optimal force estimate, let us
re-write \eqns{expand} as weight-function estimate
\begin{equation}
        \hPhi(\bx) = {\textstyle\sum}_i m_i\,W(\bx_i,\bx), \;\;
        W(\bx,\by) \equiv {\textstyle\sum}_{\bmath{n}}{\it\Phi}_{\bmath{n}}(\bx)
					    {\it\Phi}_{\bmath{n}}(\by).
\end{equation}
The effective local softening length $\epsilon$ may be defined as the distance
between zeros of the highest-order basis functions. While this varies with
position and direction, it is obvious that $\epsilon\propto n^{-1}_{\rm
max}$. The bias of the estimated force may be estimated in close analogy to the
derivation in Appendix~\ref{app:fixed:bias}. However, because $W(\bx,\bx+\bz)$
is not reflection symmetric w.r.t.\ $\bz=0$, the bias is ${\cal O}(\epsilon)$
and not ${\cal O}(\epsilon^2)$ as for kernel estimates. With ${\rm
var}_\bx(\hbF)\propto 1/N\epsilon$ we then find for the optimal softening
$n_{\rm max}\propto N^{1/3}$ and $\moptF\propto N^{-2/3}$.

Thus, the main problem of the expansion technique is its large bias, arising
from a mismatch, in all but special situations, between the stellar system and
the set of basis functions. In order to overcome this fundamental problem, one
must ensure that the low-order basis functions represent the stellar system
fairly well \cite{ho92}, for instance, by automatically adapting them to the
stellar system modelled \cite{w99}.
\subsubsection{Grid-based codes} \label{sec:disc:grid-based}
Grid-based methods for the force evaluation are also quite common. In these
methods, one estimates the density (via a kernel-estimator) on the nodes of a
regular grid, solves for the potential and forces on these nodes by some grid
method such as fast Fourier transforms, and finally interpolates the forces at
the body positions. The density errors follow the relations of
Section~\ref{sec:errors}, but do not propagate trivially to the force estimate,
because the grid method itself introduces further biases. Therefore, the
detailled results of this paper may not be directly applied to grid-based
methods, but the general remarks made at the beginning of this Section still
hold.
\subsection{Two-body relaxation}\label{sec:disc:relax}
As already explained in the introduction, softening cannot much reduce the
artificial two-body relaxation. The analytic results of Section~\ref{sec:errors}
actually support this, which can be seen as follows.

Since relaxation is is driven by the graininess of gravity, it thus must be
related to the variances of force and potential. In the limit $\epsilon\to0$ of
Newtonian gravity, the variations of the force actually diverge. However, in
order to obtain the orbital changes one has to integrate the forces along the
orbit, and this integral may no longer diverge. Actually, this integral must be
related to the variation of the potential, since $\Phi=\int\bF\cdot\,{\rm
d}\bx$. In contrast to the variance of the force, the variance of the potential
is finite for Newtonian gravity, and it derives from the contributions not only
of nearby stars, like the variance of the force, but from all stars. The same is
true for the two-body relaxation, which suggests that it is related to the
variance of the potential -- though the situation must be more complex, since
dynamical considerations (time variations) have to be taken into account.

From \eqn{var:pot}, it is clear that in order to reduce the integrated variance
of the potential substantially, we need
\begin{equation}
        \epsilon \simeq b_\Phi^{-1} \;
                {\int\dx\;\rho(\bx)\;{\rm var}_\bx^0(\Phi) \over
                 \int\dx\;\rho^2(\bx)}.
\end{equation}
For the Plummer sphere as underlying model and, say, the F$_1$ softening kernel,
we get $\epsilon\simeq1.1$ scale radii, implying that, unless the softening
length is comparable to the size of the system itself, the variance of the
potential cannot be significantly reduced by means of softening. This suggests
that the usual small-scale softening does not much reduce the two-body
relaxation in agreement with the arguments given in
Section~\ref{sec:intro:relax}.
\ifpreprint\begin{table}
 \caption{Scaling with $N$ of the $\epsilon_{\rm opt}$ and \moptF\ for some
          softening methods applied to a Hernquist model\label{tab:compare}}
 \small
 \begin{tabular}{ll@{ \& }l r@{$\,{\times}\,$}lr@{$\,{\times}\,$}l}
 kernel  & \multicolumn{2}{c}{$\alpha$}
         & \multicolumn{2}{c}{$\moptF$}
         & \multicolumn{2}{c}{$\epsilon_{\rm opt}$}	                \\ \hline
 Plummer & \multicolumn{2}{l}{0}
         & 0.000296&$N_5^{-0.77}$
         & 0.017&$N_5^{-0.23}$			\\[1ex]
 F$_1$	 & \multicolumn{2}{l}{0}
         & 0.000208&$N_5^{-4/5}$
         & 0.051 &$N_5^{-1/5}$			\\[1ex]
 K$_1$   & \multicolumn{2}{l}{0}
         & 0.000153&$N_5^{-8/9}$
         & 0.115&$N_5^{-1/9}$			\\[1ex]
 F$_1$	 & \multicolumn{2}{l}{$1/3$}
         & 0.000125&$N_5^{-4/5}$
         & 0.186&$N_5^{-1/5}$			\\[1ex]
 K$_1$   & \multicolumn{2}{l}{$1/3$}
         & 0.000102&$N_5^{-8/9}$
         & 0.309&$N_5^{-1/9}$			\\ \hline
 \end{tabular}\par\medskip
        The \moptF\ was evaluated numerically for the Hernquist model sampled
        with $N=10^5$ points (see Section~\ref{sec:exper}); $N_5\equiv N/10^5$.
        A sensitivity of $\alpha=0$ corresponds to a fixed-kernel estimate,
        i.e.\ constanst $\epsilon$.
\end{table} \fi
\subsection{What softening technique should one use?}
Let us re-consider the scaling relations of the \miseF\ and $\epsilon_{\rm opt}$
with $N$.  For various softening techniques, Table~\ref{tab:compare} lists them
in the case of $N\sim10^5$ and the Hernquist model as target. The numbers given
in this table are based on the numerical results of Section~\ref{sec:exper} and
the scaling relations of Section~\ref{sec:opt} also summarized in
Table~\ref{tab:asym} below.
\subsubsection{The softening method}
We first concentrate on simple fixed-kernel estimators. When the standard
Plummer kernel is replaced by a compact kernel (e.g.\ F$_1$) or a compensating
kernel (e.g.\ K$_1$), the \moptF\ drops by a factor of $\sim1.5$ and $\sim2$,
respectively. Note that such a change requires no overhead at all. Even codes
using the cubic spline kernel of Monaghan \& Lattanzio (1985) might benefit from
employing the F$_1$ or K$_1$ kernel instead, since their computation involves
less operations and the latter gives more accurate forces.

Another improvement by a factor of $\sim1.5$ can be achieved by adapting the
softening lengths locally in such a way that the number $N_{\rm soft}$ of bodies
in the softening volume is roughly constant (but dependent on $N$).  This result
for the Hernquist model as target applies to a lesser degree to stellar system
with a smaller dynamic range, e.g.\ with a density core like the Plummer model.
\subsubsection{How to choose $\epsilon$?}\label{sec:disc:eopt}
There are actually two questions behind this one, both of vital importance to
the subject, neither of which has been addressed so far. First, how to find the
value for $\epsilon$ that minimizes the \miseF\ if the true underlying force
field is unknown (e.g.\ during a \Nbd\ simulation) and, second, is
$\epsilon_{\rm opt}$ derived in this way really the best choice?

Merritt (1996) presumed that a time-intensive boot-strap algorithm would be
needed to find $\epsilon_{\rm opt}$ if $\bF(\bx)$ were unknown. However, if $N$
is large enough for the asymptotic relations of Section~\ref{sec:errors} to
hold, one may exploit these in order to find $\epsilon_{\rm opt}$ with an
overhead comparable to one full force computation for all bodies. The details of
this method and some performance tests will be given in a follow-up paper.

\ifpreprint\begin{table}
  \caption{Asymptotic scaling relations with $\epsilon$ and $N$ \label{tab:asym}}
  \small
  \begin{tabular}{llll}
  kernel & \multicolumn{1}{l}{${\rm bias}(\hbF)$}
	 & \multicolumn{1}{c}{$\epsilon_{\rm opt}$}
	 & \multicolumn{1}{c}{$\moptF$}    		\\ \hline
  Plummer     &$\propto\epsilon^{1.67}$&$\propto N^{-0.23}$&$\propto N^{-0.77}$\\
  compact     &$\propto\epsilon^{2}$   &$\propto N^{-1/5} $&$\propto N^{-4/5}$ \\
  compensating&$\propto\epsilon^{4}$   &$\propto N^{-1/9} $&$\propto N^{-8/9}$ \\
  \hline
  \end{tabular}\par\medskip
  Note that ${\rm var}(\hbF)$ always scales as $\epsilon^{-1}N^{-1}$, so the
  scaling of ${\rm bias}(\hbF)$ makes all the difference.
\end{table}\fi
The second question is less technical and more difficult to answer. The error of
the force comes in two parts, bias and variance. In the time mapping of the DF,
these will give rise to different artificial effects. The variance results in
enhanced relaxation, while the bias modifies the stellar dynamics; moreover
there might be some interplay between the two. It is not clear and non-trivial
to say, what the consequences of these modifications are and how they scale with
simulation time. It is likely that the optimal $\epsilon$ depends both on the
stellar system modelled and on the specific goal of the simulation. If, for
instance, the stability of some equilibrium configuration is to be investigated,
it is essential to modify gravity as little as possible, i.e. use small
$\epsilon$ (and a compensating kernel)\footnote{ In these circumstances, a
technique other than softening can and should be used for variance reduction:
the quiet start \cite{se83}. In this method, the initial phase-space positions
are setup essentially noise free and, {\em if} the motion is predominantly
regular, remain so for several dynamical times.}.  On the other hand, when more
violent dynamics is investigated (e.g.~mergers) one might want to suppress
small-scale noise. Obviously, these are important issues which deserve further
investigations.

\subsection{The optimal collisionless $\bmath{N}$-body code}
Merritt (1996) proposed to call a Poisson solver optimal if it results in the
smallest \miseF\ for given $N$. With this definition, all non-direct methods are
sub-optimal (since, in addition to the softening, they invoke approximations,
which introduce additional errors), and yet such methods are generally
preferred. Therefore, one may better define: {\em The optimal collisionless
$N$-body code achieves, for given computer resources (CPU time and memory), the
most faithful time-mapping of those properties of the DF that are essential for
the purpose of the particular simulation}. Note that in this definition, the
number $N$ of bodies does not occur, $N$ is just a parameter of the code, as is
the softening length. This means in particular, that contrary to widespread
practice, the number $N$ alone is not a sufficient criterion for the quality of
a \Nbd\ simulation.

What is the optimal Poisson solver for general-purpose collisionless \Nbd\
codes? From the discussion in Section~\ref{sec:disc:relev}, it is clear that
expansion techniques are not suitable for a general applicable \Nbd\ code --
unless, perhaps, Weinberg's (1999) automatic adaption method is continuously
applied. Obviously, the optimal Poisson solver must be adaptive in space and
time. This condition clearly favours the tree code, which naturally is fully
adaptive, while for grid-based solvers a considerable effort is needed to make
them fully adaptive.

From the results of this paper, it appears to be obvious that the optimal
softening method uses adaptive softening lengths and a compact, or even
compensating, kernel. A fixed softening length may be appropriate in the case of
a stellar system with a small dynamic range, i.e.\ with a constant-density core.

\subsection{On the resolution of a $\bmath{N}$-body simulation}\label{sec:resolv}
In discussing \Nbd\ simulations, it is important to know their resolution scale,
i.e.\ the smallest scale on which a simulation is ``believable''. Lacking a more
precise definition, one often simply uses the softening length $\epsilon$ for
the resolution scale. With the asymptotic relations in Table~\ref{tab:asym},
this yields the absurd result that users of Plummer softening can improve the
resolution of their simulations faster with increasing $N$ than users of compact
or compensating kernels, which we have shown to be superior to Plummer
softening.  How can this be?

The problem is that linearly relating $\epsilon$ with resolution scale is
incorrect. The softening length $\epsilon$ is a mere parameter without any
physical meaning. A force resolution length may be properly defined as the scale
over which the maximum change in the true force equals the (local) error of the
estimated force:
\begin{equation}\label{res}
	s_{\rm res}(\bx) \approx {\sqrt{{\rm MSE}_\bx(\hbF)}\over
			\|\bnabla\otimes \bF(\bx)\|}.
\end{equation}
With the results of Section~\ref{sec:errors}, we find that, for optimally chosen
$\epsilon$, the decrease of $s_{\rm res}$ with increasing $N$ is slowest for
Plummer softening and fastest for softening with compensating kernels. In
particular, the relation between $\epsilon$ and $s_{\rm res}$ is
non-linear\footnote{ Note, however, that in a cusp $\bF$ becomes discontinuous
and \eqn{res} invalid. In such a situation one has indeed $s_{\rm
res}\sim\epsilon$.}.

Note, however, that $s_{\rm res}$ defined above is a mere measure for the
resolution of the {\em force}, and not of the believability of the
simulation. Since gravity is a long-range force, local errors in the DF,
introduced on scales of $s_{\rm res}$, may well propagate to larger scales.

\section{Summary and conclusion}\label{sec:summ}
In collisionless \Nbd\ simulations, the bodies are the numerical representation
of the continuous one-particle DF. Because $N$ is much smaller than in the
stellar system being modelled, this representation is always incomplete, i.e.\
noisy. In order to estimate, at each time step, the continuous force field due
to the DF, one has to moderate this noise. This exactly is the purpose of force
softening in collisionless \Nbd\ codes. However, while reducing the noise in the
\Nbd\ forces, the softening modifies the laws of gravity: the force vanishes at
vanishing inter-body separation $r\to0$. This generates a bias, a systematic
offset between the estimated and true forces. Both, the noise and the bias,
contribute to the mean square error of the estimated force field. While the
noise decreases with the softening length $\epsilon$, the bias increases, and
the optimal softening represents a compromise between these two
\cite{merritt96}.

In the limit of $\epsilon$ being small compared to any physical scale of the
stellar system, a requirement that is desirably satisfied in any \Nbd\
simulation, analytic expressions for the noise-generated variance of the force
and for the bias have been derived. While the relations \eqb{bias} for the bias
have already been given in the literature (e.g.\ Hernquist \& Katz 1989), those
for the variance of the estimated potential and force \eqi{vars} seem to be
hitherto unknown. One of the most important results is that the variance of the
force is a local quantity diverging in the limit $\epsilon\to0$, very much like
the variance of the density, while the variance of the potential depends on a
global measure and reaches a finite value for $\epsilon=0$. Presumably, this is
related to the fact, see Section~\ref{sec:disc:relax} and Theis (1998), that the
two-body relaxation is not substantially reduced by force softening.

In these asymptotic relations, the effects of the softening length, the
softening kernel, the number $N$ of bodies, and of the underlying stellar system
nicely separate, allowing simple and, in the important case of large $N$,
accurate estimates for the behaviour of the mean integrated squared force error,
\miseF.  For the various types of softening kernels, Table~\ref{tab:asym} gives
the asymptotic relations for the optimal softening length and force errors (the
relations given for Plummer softening are based on numerical experiments
only). The strongest possible dependence of the MISE on $N$ is $N^{-1}$ as
pertains, e.g., to parametric estimators.  Note that the compensating kernels
come very close to this limit. The main results of this investigation are:
\setcounter{lll}{0}
\begin{list}{\arabic{lll}.}{\usecounter{lll} \leftmargin4mm 
\itemsep0pt plus 1pt
\parsep0pt plus 1pt
\topsep2pt plus 1pt
\labelwidth3mm \labelsep1mm \itemsep2ex}
\item   Compact softening kernels, i.e.\ those with finite density support, are
        superior to the standard Plummer softening. This improvement is ever
        larger for increasing $N$.
\item   Special softening kernels are derived, for which the force errors are
        even smaller. These kernels compensate the small forces at $r\to0$ by
        forces larger than Newtonian at $r\sim\epsilon$.
\item   For inhomogeneous stellar systems, such as galaxies, adaptive softening
        lengths allow a further substantial reduction of the force errors.
\end{list}
All these numbers are based on the Hernquist model, which gives a fair
representation of a typical galaxy, as underlying stellar system, resulting in a
rms force error of $\sim0.01=0.04$ of the rms force of this model. For smaller
errors to be achieved, the advantage of using compensating kernels and adaptive
softening lengths will become even larger.

For this error of 4\%, particle numbers $N\sim10^5$ are necessary, and roughly
20 times as many to reduce the error to 1\%. It is not clear, what these errors
imply in terms of the reliability of the time evolution of a corresponding \Nbd\
simulation. This is a very important question to answer, in order to better
understand the results of \Nbd\ simulations and assess their reliability. For
example, current \Nbd\ simulations of large-scale structure formation employ
only about $10^{3-4}$ bodies per halo, which come out to have a profile similar
to a Hernquist model, i.e.\ the rms force error is $10\%$ to $25\%$, even if
$\epsilon$ was chosen optimally, which it very likely was not.

\section*{Acknowledgements}
The authors thanks the referee, Joshua Barnes, as well as David Merritt, Jerry
Sellwood, James Binney, Rainer Spurzem, and Daniel Pfenniger for many helpful
comments and discussions.

\ifpreprint \relax \else
\begin{table*}
        \caption{Some 3D softening kernels and their properties
                \label{tab:kernels}}
        \footnotesize
	\begin{tabular}{llcccccccr}
        $\!\!$name                                          &
        \hspace{4em}{kernel density $\eta(r)$}              &
        \multicolumn{1}{c}{$r_{{\rm max}\,F}$}              &
        \multicolumn{1}{c}{$\phi^\prime_{\rm max}$}            &
        \multicolumn{1}{c}{$a_0$}                           &
        \multicolumn{1}{c}{$a_2$}                           &
        \multicolumn{1}{c}{$b_\Phi$}                        &
        \multicolumn{1}{c}{$b_F$}                           &
        \multicolumn{1}{c}{$b_\rho$}                        &
        \multicolumn{1}{c}{$E_F$}                           \\  \hline
        $\!\!$Plummer
                & $\displaystyle{3\over4\pi}{1\over(1+r^2)^{5/2}}$
		& 0.707107 & 0.384900
                & $\infty$
                & $\infty$
                & $2\pi^2$
                & $\displaystyle{3\pi^2\over4}$
                & $\displaystyle{45\over1024}$
                & \multicolumn{1}{c}{--}
\\[4ex]	$\!\!$cubic spline  
                & $\left\{\begin{array}{l} {1\over4\pi}(4-6r^2+3r^3)    \\[1ex]
                           {1\over4\pi}(2-r)^3          \\[1ex]
                            0   \end{array}\right.$ \hfill
                  $\begin{array}{r} r<1 \\[1ex] 1\le r<2 \\[1ex] r\ge2
		   \end{array}$
		& 0.828302 & 0.657817
                & $\displaystyle{3\pi\over5}$
                & $\displaystyle{17\pi\over60}$
                & $\displaystyle{1120789\pi\over450450}$
                & $\displaystyle{70016\pi\over17325}$
                & $\displaystyle{491\over1260\pi}$
                & 9.84
\\[6ex]	$\!\!$F$_0$
		& $\displaystyle{3\over4\pi}\,H(1-r^2)$
		& 1 & 1
                & $\displaystyle{2\pi\over5}$
                & $\displaystyle{\pi\over70}$
                & $\displaystyle{72\pi\over35}$
                & $\displaystyle{24\pi\over5}$
                & $\displaystyle{3\over4\pi}$
                & 9.60
\\[3ex]	$\!\!$F$_1$
                & $\displaystyle{15\over8\pi}(1-r^2)\,H(1-r^2)$
		& 0.745356 & 1.242260
                & $\displaystyle{2\pi\over7}$ 
                & $\displaystyle{\pi\over126}$ 
                & $\displaystyle{400\pi\over231}$
                & $\displaystyle{40\pi\over7}$
                & $\displaystyle{15\over14\pi}$
                & 9.65
\\[3ex]$\!\!$F$_2$
                & $\displaystyle{105\over32\pi}(1-r^2)^2\,H(1-r^2)$
		& 0.592614 & 1.637096
                & $\displaystyle{2\pi\over9}$ 
                & $\displaystyle{\pi\over198}$ 
                & $\displaystyle{1960\pi\over1287}$
                & $\displaystyle{2800\pi\over429}$
                & $\displaystyle{35\over22\pi}$
                & 9.71
\\[3ex]$\!\!$F$_3$
                & $\displaystyle{315\over64\pi}(1-r^2)^3\,H(1-r^2)$
		& 0.505871 & 2.051564
                & $\displaystyle{2\pi\over11}$ 
                & $\displaystyle{\pi\over186}$ 
                & $\displaystyle{63504\pi\over46189}$
                & $\displaystyle{17640\pi\over2431}$
                & $\displaystyle{315\over143\pi}$
                & 9.75
\\[4ex] $\!\!$K$_0$   
                & $\displaystyle{15\over16\pi}(5-7r^2)\,H(1-r^2)$
		& 0.629941 & 2.624753
                & 0
                & $\displaystyle-{\pi\over126}$
                & $\displaystyle{170\pi\over231}$
                & $\displaystyle{10\pi}$
                & $\displaystyle{75\over16\pi}$
                & 9.43
\\[3ex]$\!\!$K$_1$   
                & $\displaystyle{105\over64\pi}(5-9r^2)(1-r^2)\,H(1-r^2)$
		& 0.493924 & 3.436176
                & 0
                & $\displaystyle-{\pi\over198}$ 
                & $\displaystyle{280\pi\over429}$
                & $\displaystyle{1610\pi\over143}$
                & $\displaystyle{525\over88\pi}$
                & 9.48
\\[3ex]$\!\!$K$_2$   
                & $\displaystyle{315\over128\pi}(5\,{-}\,11r^2)(1\,{-}\,r^2)^2\,H(1-r^2)$
		& 0.419491 & 4.296037
                & 0
                & $\displaystyle-{\pi\over286}$ 
                & $\displaystyle{27342\pi\over46189}$
                & $\displaystyle{30240\pi\over2431}$
                & $\displaystyle{9135\over1144\pi}$
                & 9.54
\\[3ex]$\!\!$K$_3$   
                & $\displaystyle{3465\over1024\pi}(5\,{-}\,13r^2)(1\,{-}\,r^2)^3
			\,H(1-r^2)$
		& 0.370935 & 5.170685
                & 0
                & $\displaystyle-{\pi\over390}$ 
                & $\displaystyle{2772\pi\over5083}$
                & $\displaystyle{56826\pi\over4199}$
                & $\displaystyle{86625\over8398\pi}$
                & 9.60
\\ \hline
\end{tabular}\par\medskip\begin{minipage}{175mm}
	The kernels F$_n$ (Ferrers (1877) spheres of index $n$) and K$_n$ are
	continuous in the first $n$ force derivatives ($H$ denotes the Heaviside
	function). The potential of the Plummer kernel is $\phi(r)=1/
	\sqrt{1+r^2}$, that of the cubic spline kernel is given by Hernquist \&
	Katz (1989), while for the kernels F$_n$ and K$_n$, the potentials are
	low-order polynomials in $r^2$ and can be found in
	Appendix~\ref{app:kernels}.  $r_{{\rm max}\,F}$ is the radius, in units
	of $\epsilon$, of the maximal force, while $\phi^\prime_{\rm
	max}\equiv\phi^\prime( r_{{\rm max}\,F})$. The constants $a_0$ and $a_2$
	are related to the bias introduced by the softening \eqis{bias}, while
	the constants $b_\Phi$, $b_F$, and $b_\rho$ are related to the variances
	\eqis{vars}.  $E_F$ (defined in the text) is a measure for the
	efficiency of the kernel: for fixed $N$ and optimal $\epsilon$, the mean
	integrated squared force error, \miseF, is directly proportional to
	$E_F$, though with a different constant of proportionality for compact
	non-negative kernels (spline and F$_n$) and compensating kernels
	(K$_n$), respectively.  \end{minipage}
\vspace*{10cm}
\end{table*}
\begin{table}
 \caption{Scaling with $N$ of the $\epsilon_{\rm opt}$ and \moptF\ for some
          softening methods applied to a Hernquist model\label{tab:compare}}
 \small
 \begin{tabular}{ll@{ \& }l r@{$\,{\times}\,$}lr@{$\,{\times}\,$}l}
 kernel  & \multicolumn{2}{c}{$\alpha$}
         & \multicolumn{2}{c}{$\moptF$}
         & \multicolumn{2}{c}{$\epsilon_{\rm opt}$}	                \\ \hline
 Plummer & \multicolumn{2}{l}{0}
         & 0.000296&$N_5^{-0.77}$
         & 0.017&$N_5^{-0.23}$			\\[1ex]
 F$_1$	 & \multicolumn{2}{l}{0}
         & 0.000208&$N_5^{-4/5}$
         & 0.051 &$N_5^{-1/5}$			\\[1ex]
 K$_1$   & \multicolumn{2}{l}{0}
         & 0.000153&$N_5^{-8/9}$
         & 0.115&$N_5^{-1/9}$			\\[1ex]
 F$_1$	 & \multicolumn{2}{l}{$1/3$}
         & 0.000125&$N_5^{-4/5}$
         & 0.186&$N_5^{-1/5}$			\\[1ex]
 K$_1$   & \multicolumn{2}{l}{$1/3$}
         & 0.000102&$N_5^{-8/9}$
         & 0.309&$N_5^{-1/9}$			\\ \hline
 \end{tabular}\par\medskip
        The \moptF\ was evaluated numerically for the Hernquist model sampled
        with $N=10^5$ points (see Section~\ref{sec:exper}); $N_5\equiv N/10^5$.
        A sensitivity of $\alpha=0$ corresponds to a fixed-kernel estimate,
        i.e.\ constanst $\epsilon$.
\end{table}
\begin{table}
  \caption{Asymptotic scaling relations with $\epsilon$ and $N$ \label{tab:asym}}
  \small
  \begin{tabular}{llll}
  kernel & \multicolumn{1}{l}{${\rm bias}(\hbF)$}
	 & \multicolumn{1}{c}{$\epsilon_{\rm opt}$}
	 & \multicolumn{1}{c}{$\moptF$}    		\\ \hline
  Plummer     &$\propto\epsilon^{1.67}$&$\propto N^{-0.23}$&$\propto N^{-0.77}$\\
  compact     &$\propto\epsilon^{2}$   &$\propto N^{-1/5} $&$\propto N^{-4/5}$ \\
  compensating&$\propto\epsilon^{4}$   &$\propto N^{-1/9} $&$\propto N^{-8/9}$ \\
  \hline
  \end{tabular}\par\medskip
  Note that ${\rm var}(\hbF)$ always scales as $\epsilon^{-1}N^{-1}$, so the
  scaling of ${\rm bias}(\hbF)$ makes all the difference.
\end{table}
\fi 
\onecolumn \appendix
\section{Asymptotic Relations for Bias and Variance}\label{app:asym}
\subsection{Fixed-kernel estimates with equal weights}\label{app:fixed}
\subsubsection{The bias}\label{app:fixed:bias}
In calculating the expectation value, we may replace the sum over points by an
integral weighted with the density $\rho$. Thus
\begin{equation} \label{app:exp:pot:a}
        \big\langle\hPhi(\bx)\big\rangle
                        = -{G\over\epsilon}\int\dy\;\rho(\by)\;
                        \phi\left[{|\bx-\by|\over\epsilon}\right].
\end{equation}
We proceed by writing
\begin{equation}\label{app:decompose:phi}
        {1\over\epsilon}\,\phi\left({r\over\epsilon}\right)
        = {1\over r} - {1\over\epsilon} \left[ {\epsilon\over r} - 
                        \phi\left({r\over\epsilon}\right) \right].
\end{equation}
Inserting this into \eqn{app:exp:pot:a} and substituting $\by=\bx-\epsilon\bz$,
we find
\begin{equation} \label{app:exp:pot:b}
        \big\langle\hPhi(\bx)\big\rangle = \Phi(\bx) + G\,\epsilon^2\int\dz\;
                \left[{1\over|\bz|}-\phi(|\bz|)\right]\;\rho(\bx-\epsilon\bz).
\end{equation}
When replacing $\rho(\bx-\epsilon\bz)$ by its Taylor expansion about $\bx$,
\begin{equation} \label{rho:Taylor}
        \rho(\bx-\epsilon\bz) = \sum_{k\ge0}{(-\epsilon)^k\over k!}
                        (\bz\bnabla)^k\rho(\bx),
\end{equation}
we see that, by virtue of the symmetry of the kernel, the integrals over all 
odd orders in $\epsilon$ vanish identically and one obtains
\begin{equation}
        {\rm bias}_\bx(\hPhi) = a_0\,\epsilon^2G\,\rho(\bx)
                + a_2\,\epsilon^4G\,\Delta\rho(\bx) + {\cal O}(\epsilon^6)
\end{equation}
where
\ifpreprint
\begin{equation} \label{a}
        a_k     = {4\pi\over(k+1)!} \int_0^\infty{\rm d}r\;r^{k+2}
                        \left[{1\over r}-\phi(r)\right]
                = {4\pi\over(k+1)!\,(k+3)} \int_0^\infty{\rm d}r\;r^{k+3}
                        \left[{1\over r^2}+\phi^\prime(r)\right]
                = {(4\pi)^2\over(k+3)!} \int_0^\infty{\rm d}r\;r^{k+4}
                        \eta(r).
\end{equation}
\else
\begin{eqnarray}
        a_k     &=& {4\pi\over(k+1)!} \int_0^\infty{\rm d}r\;r^{k+2}
                        \left[{1\over r}-\phi(r)\right]	\nonumber	\\
                &=& {4\pi\over(k+1)!\,(k+3)} \int_0^\infty{\rm d}r\;r^{k+3}
                        \left[{1\over r^2}+\phi^\prime(r)\right]		\\
                &=& {(4\pi)^2\over(k+3)!} \int_0^\infty{\rm d}r\;r^{k+4}
                        \eta(r).\nonumber
\end{eqnarray}
\fi
A similar estimate can be made for the bias of the force and density estimators
in \eqns{soft:force} and \eqb{soft:density}, but it is actually simpler to
relate the biases directly using Poisson's equation. Note, that the integrals
over $\dz$ in \eqn{app:exp:pot:b} may not converge if the kernel does not
approach its asymptotic shape $\phi\to 1/r$ fast enough at large radii. To be
more precise, if the density kernel $\eta$ falls off as $r^{-5}$ or less
steeply, then the biases grow faster than quadratically with small $\epsilon$.

Furthermore, if the Taylor expansion in \eqn{rho:Taylor} does not converge, the
above estimate is incorrect, too, usually over-estimating the true bias. To see
this, consider a spherical cusp in the density at $\by=0$, i.e.\ $\rho=\rho_0
y^{-\gamma}$ (with $\gamma<2$) at small $y$. Inserting this instead of the
Taylor expansion into \eqn{app:exp:pot:b}, we find for $x\ll\epsilon$
\begin{subequations} \label{bias:cusp} \begin{eqnarray}
        {\rm bias}_{\rm cusp}(\hPhi) &\approx&
	G\,\rho_0\,\epsilon^{2-\gamma}\,4\pi
        \int_0^\infty{\rm d}r\;r^{2-\gamma}\left[{1\over r}-\phi(r)\right]
\\[1ex]
        {\rm bias}_{\rm cusp}(\hbF) &=&
	-G\,\rho_0\,\gamma\,\epsilon^{1-\gamma}\,4\pi
        \int_0^\infty{\rm d}r\;r^{1-\gamma}\left[{1\over r}-\phi(r)\right].
\end{eqnarray} \end{subequations}
Here, the integrals extend only formally to infinity: for a sensible kernel, the
term in brackets vanishes outside some small radius. Thus, for cusps shallower
than $\gamma=1$, the bias in the force is still finite; for steeper cusps, the
correct force actually diverges, which of course cannot be reproduced by the
softening, resulting in a diverging force bias.
\subsubsection{The variance} \label{app:fixed:var}
Assuming the positions $\bx_i$ are independent, the variance of some quantity is
equal to $N^{-1}$ times the variance of the contribution of one point. Moreover,
the central-limit theorem tells us that the estimates follow essentially a
normal distribution.
\nobreak
\paragraph*{The variance of the estimated potential.}
With $\br_i=\bx_i-\bx$, the contribution of one point is
$\phi_i(\bx)\equiv-GM\,\epsilon^{-1}\phi(|\br_i|/\epsilon)$. Thus, from
\eqn{var} we have
\begin{equation}
        N {\rm var}_\bx(\hPhi) = \big\langle\phi_i^2(\bx)\big\rangle
                                - \big[\Phi(\bx)+{\rm bias}_\bx(\hPhi)\big]^2.
\end{equation}
We may evaluate $\langle\phi\rangle_i^2$ as an average weighted with the density
$\rho$:
\begin{equation} \label{app:var:pot:a}
        \big\langle\phi_i^2(\bx)\big\rangle 
        =       {G^2M\over\epsilon^2} \int\dy\;\rho(\by)\,
                \phi^2\left({|\bx-\by|\over\epsilon}\right).
\end{equation}    
We proceed by writing
\begin{equation}
        {1\over\epsilon^2}\;\phi^2\Big({r\over\epsilon}\Big)=
                {1\over r^2}+{1\over\epsilon^2}\left
                [\phi^2\Big({r\over\epsilon}\Big)-{\epsilon^2\over r^2}\right]
\end{equation}
to finally obtain
\begin{equation}
        \big\langle\phi^2_i(\bx)\big\rangle = G^2M\left\{
		\left[\int\dy\;{\rho(\by)\over(\bx-\by)^2}\right]
                - \epsilon\,b_\Phi\,\rho(\bx) + {\cal O}(\epsilon^2)
		\right\}
\end{equation}
with $b_\Phi = 4\pi\,\int_0^\infty{\rm d}r\,[1-r^2\phi^2(r)]$.
\paragraph*{The variance of the estimated force.}
The contribution from one body is
$\bmath{f}_i=\br_i/|\br_i|\,G\,M\,\epsilon^{-2}\phi^\prime(|\br_i|/\epsilon)$,
and in analogy to \eqn{app:var:pot:a}, we have
\begin{equation}\label{app:var:for:a}
        \big\langle\bmath{f}_i\otimes\bmath{f}_i\big\rangle =
		{G^2M\over\epsilon^4}\int\dy\;
                \rho(\by)\;\phi^{\prime2}\left({|\bx-\by|\over\epsilon}\right)\;
                {(\bx-\by)\otimes(\bx-\by)\over(\bx-\by)^2}.
\end{equation}
Substituting $\by=\bx-\epsilon\bz$ and Taylor expanding $\rho$, we find for the
variance
\begin{equation} \label{app:var:for}
	{\rm var}_\bx(\hbF) = G^2\,\epsilon^{-1}\,M\,N^{-1}\,b_F\,
	\rho(\bx)\,{\textstyle{1\over3}}\bmath{\sf I}\;+\;{\cal O}(\epsilon^0),
\end{equation}
where $\bmath{\sf I}$ denotes the unit matrix, while
$b_F=4\pi\,\int_0^\infty{\rm d} r\,r^2\phi^{\prime2}(r)$.
\paragraph*{The variance of the estimated density.}
The contribution from one body is $\rho_i=M\,\epsilon^{-3}\eta(|\br_i|
/\epsilon)$, and 
\begin{equation}
        \big\langle\rho_i^2\big\rangle =
		{M^2 \over\epsilon^6} \int\dy\;\rho(\by)\;
                \eta^2\left({|\bx-\by|\over\epsilon}\right)
                  = M^2\epsilon^{-3}\,b_\rho\,\rho(\bx) + {\cal O}(\epsilon^{-1})
\end{equation}
with $b_\rho = \int_0^\infty{\rm d}r\,r^2\eta^2(r)$.
\subsection{Adaptive softening with equal weights}\label{app:adap}
In analogy to \eqn{app:exp:pot:b}, we get for the expectation value of the
estimated potential
\begin{equation} \label{app:bias:adap:pot:a}
        \big\langle\hPhi(\bx)\big\rangle = \Phi(\bx) + G\epsilon^2 \int\dz\;
                {\rho(\bx-\epsilon\bz)\over\lambda}\,
                \left[{1\over r}-\phi(r)\right]_{r=|\bz|/\lambda}.
\end{equation}
For equal weights, \eqn{lambda} reduces to $\lambda=[\hrho(\bx-\epsilon\bz)
/\overline{\rho}]^{-\alpha}$. In order to estimate the asymptotic behaviour at
small $\epsilon$, we may replace $\hrho$ in this relation by $\rho$. For kernels
with $a_0\neq0$, the error made by this assumption is ${\cal O}(\epsilon^2)$ and
hence the resulting error in the bias of the potential is a factor $\epsilon^2$
smaller than the leading term. We thus get
\begin{equation} \label{app:adap:bias:pot}
        {\rm bias}_\bx(\hPhi) = a_0\,G\,\epsilon^2\,\overline{\rho}^{2\alpha}
                        \rho(\bx)^{1-2\alpha}+{\cal O}(\epsilon^4).
\end{equation}
which for $\alpha=0$ reduces to the result for the fixed-kernel estimate.

For $\alpha=1/2$, the quadratic term in ${\rm bias}(\hPhi)$ is constant and the
biases of force and density are formally only ${\cal O}(\epsilon^4)$. However,
this was not anticipated in the approximation made above, where $\hrho$ in the
definition of $\lambda$ was replaced by $\rho$. Thus, for this special case, the
estimation might be incorrect.

Because the variances are dominated by the lowest order in $\epsilon$, we may use
the above relations for the fixed-kernel estimates with $\epsilon$ replaced with
$\epsilon[\rho(\bx)/\overline{\rho}]^{-\alpha}$.
\subsection{Softening with individual weights}\label{app:ind}
In this case, the softening lengths $\epsilon\lambda_i$ depend on the full 6D
phase-space coordinates, and we must replace the density-weighted averaging over
configuration space by a sampling-distribution-function-weighted averaging over
phase space.
\subsubsection{The bias}\label{app:ind:bias}
With \eqn{app:decompose:phi} we obtain
\begin{equation} \label{app:exp:pot:w}
        {\rm bias}_\bx(\hPhi) = {G\over\epsilon}\int\dy\int\dv\;f_{\rm s}(\by,\bv)
        \;{\mu(\by,\bv)\over\lambda(\by,\bv)}\; \left[{1\over r}-\phi(r)
        \right]_{r=|\bx-\by|/\epsilon\lambda(\by,\bv)}
\end{equation}
with
\begin{equation} \label{app:lambda}
	\lambda(\by,\bv)=\mu(\by,\bv)^{1/2}\;[\hn(\by)/\overline{n}]^{-\alpha},
\end{equation}
where $\mu(\by,\bv)$ is given in \eqn{mu}. In order to proceed, we first replace
$\hn$ by $n$ in the definition of $\lambda$ in analogy to the equal-weight case,
and second substitute $\by=\bx-\epsilon\lambda\bz$ with $|{\rm d}\by/{\rm
d}\bz|=\epsilon^3\lambda^3 (\bx,\bv)+{\cal O}(\epsilon^4z)+{\cal
O}(\epsilon^5z^2)$. We then finally get
\begin{equation}
        {\rm bias}_\bx(\hPhi) =
		a_0\,G \epsilon^2\,\rho(\bx)\,\overline{\mu}(\bx)\,
        [n(\bx)/\overline{n}]^{-2\alpha} \;+\;{\cal O}(\epsilon^4),
\end{equation}
where $\overline{\mu}=\overline{\mu^1}$ with $\overline{\mu^a}\equiv\rho^{-1}
\int\dv f \mu^a$ the mass-weighted average of $\mu^a$ of all trajectories
passing through $\bx$.
\subsubsection{The variance}\label{app:ind:varss}
For the force variance, we have instead of \eqn{app:var:for:a}
\begin{equation}
        \big\langle\bmath{f}_i\otimes\bmath{f}_i\big\rangle =
		{G^2M\over\epsilon^4}\int\dy\int\dv\;
		f_{\rm s}(\by,\bv)\,{\mu(\by,\bv)^2\over\lambda(\by,\bv)^4}\;
                \phi^{\prime2}\left({|\bx-\by|\over\epsilon}\right)\;
                {(\bx-\by)\otimes(\bx-\by)\over(\bx-\by)^2}.
\end{equation}
When substituting $\by=\bx-\epsilon\lambda\bz$ and using \eqn{app:lambda}, we get
for the force variance to lowest order in $\epsilon$
\begin{equation}
        {\rm var}_\bx(\hbF) = G^2\,\epsilon^{-1}\,M\,N^{-1}\,b_F\,
	\rho(\bx)\;\overline{\mu^{1/2}}(\bx)\;[n(\bx)/\overline{n}]^\alpha
	\; + {\cal O}(\epsilon^0).
\end{equation}
Note that for $\alpha=0$ these relations do not reduce to those given for
fixed-kernel estimates, because, with our definition of $\lambda_i$
\eqi{lambda}, individual weights also result in individual bandwidths.
\section{Investigations for better kernels} \label{app:kernels}
We want to design a spherical kernel $\phi(r)$ that corresponds to a compact
density $\eta(r)$ and satisfies all conditions listed in
Section~\ref{sec:kernel:opt}. Let us therefore assume that $\eta(r)=0$ and
$\phi(r)=r^{-1}$ for $r>1$. Moreover, continuity of $\phi$ and the normalization
of $\eta$ give us the constraints
\begin{equation} \label{constraints}
        \phi^\prime(0) = 0,\qquad
        \phi(1)        = 1,\qquad
        \phi^\prime(1) =-1.
\end{equation}
Let us now make an ansatz for $\phi$ that, in order to satisfy point 4 of the
list in Section~\ref{sec:kernel:opt}, involves only powers of $r^2$ for $r<1$:
\begin{equation} \label{phi:ansatz}
        \phi = \sum_{k\ge0}\,C_k\,(1-r^2)^k\quad{\rm for}\;r\le1
                \qquad{\rm and}\qquad\phi=r^{-1}\quad{\rm otherwise}.
\end{equation}
The constraints \eqb{constraints} require $C_0=1$ and
$C_1={\textstyle{1\over2}}$, but do not require anything for the higher-order
coefficients. When these are all set to zero, one obtains the homogeneous-sphere
kernel.

\subsection{Ferrers sphere kernels}
The kernel of the form \eqb{phi:ansatz} with
\begin{equation} \label{Ck}
	C_0 = 1,\qquad
	C_1 = {\textstyle{1\over2}},\qquad
        C_k = {1\over2^{2k-1}}{2k-1\;\choose k}\quad{\rm for}\;\;k=2,\dots,n+1
\end{equation}
is continuous in the first $n$ force derivatives, or, equivalently, in the first
$(n-1)$ derivatives of the density. If in addition $C_k=0$ for $k>n+1$, we
obtain kernels that are Ferrers (1877) spheres of index $n$, denoted F$_n$ in
this paper. They have density
\begin{equation} \label{Fn:rho}
	\eta(r) = {(2n+3)!\over n!\,(n+1)!\,2^{2n+3}\,\pi}\;
		(1-r^2)^n \;H(1-r^2),
\end{equation}
where $H$ denotes the Heaviside function.  The first few are also known as
homogeneous sphere ($n=0$), Epanechnikov ($n=1)$ and biweight ($n=2$) kernel
and are listed in Table~\ref{tab:kernels}
\subsection{Compensating kernels}
Kernels with $a_0=0$ \eqi{ak}, i.e.\ with creating reduced bias, and continuity
in the first $n$ force derivatives can be obtained by setting $C_k$ for $k\ge
n+1$ as in \eqn{Ck} and
\begin{equation}
	C_{n+2} = {(2n+7)!\over(n+2)!\,(n+3)!\,(2n+5)\,2^{2n+6}},
	\quad C_k=0\quad{\rm for}\;\;k>n+2.
\end{equation}
These kernels, which are denoted K$_n$ in this paper, may be considered as
Ferrers spheres with a higher-order addition. They have density
\begin{equation} \label{Kn:rho}
	\eta(r) = {(2n+5)!\over (n+1)!\,(n+2)!\,2^{2n+6}\,\pi}\,
		(5-[2n+7]r^2)\,(1-r^2)^n\;H(1-r^2).
\end{equation}
The properties of the first four kernels K$_n$ are listed in
Table~\ref{tab:kernels}.

\ifpreprint
	\end{document}
\fi
\newpage
        \centerline{\epsfxsize=130mm \epsfbox[28 158 444 717]{MA90.fig1.eps}}
        \centerline{Figure~\ref{fig:kernels} }
\newpage
        \centerline{\epsfxsize=130mm \epsfbox[28 158 444 717]{MA90.fig2.eps}}
        \centerline{Figure~\ref{fig:kernels:impr} }
\newpage
        \centerline{\epsfxsize=110mm \epsfbox[39 157 313 715]{MA90.fig3.eps}}
        \centerline{Figure~\ref{fig:mise:P:fix} }
\newpage
        \centerline{\epsfxsize=110mm \epsfbox[39 157 313 715]{MA90.fig4.eps}}
        \centerline{Figure~\ref{fig:mise:H:fix} }
\newpage
        \centerline{\epsfxsize=110mm \epsfbox[39 157 313 715]{MA90.fig5.eps}}
        \centerline{Figure~\ref{fig:mise:P:ada:ep} }
\newpage
        \centerline{\epsfxsize=110mm \epsfbox[39 157 313 715]{MA90.fig6.eps}}
        \centerline{Figure~\ref{fig:mise:P:ada:k} }
\newpage
        \centerline{\epsfxsize=110mm \epsfbox[39 157 313 715]{MA90.fig7.eps}}
        \centerline{Figure~\ref{fig:mise:H:ada:ep} }
\newpage
        \centerline{\epsfxsize=110mm \epsfbox[39 157 313 715]{MA90.fig8.eps}}
        \centerline{Figure~\ref{fig:mise:H:ada:k} }
\newpage
\end{document}